\documentclass[longauth]{aa} 

\usepackage{graphicx}
\usepackage{rotating}
\usepackage{txfonts}
\usepackage{xcolor}
\bibpunct{(}{)}{;}{a}{}{,} 
{
\DeclareMathOperator*{\argmin}{argmin}
\begin{document}

\title{Assessing the impact of two independent direction-dependent calibration algorithms on the LOFAR 21-cm signal power spectrum}
\subtitle{And applications to an observation of a field flanking the North Celestial Pole}
\titlerunning{Assessing two direction-dependent calibration algorithms on LOFAR 21-cm signal data}


   \author{H. Gan\inst{1}\and
        F. G. Mertens \inst{2,1}\and 
        L. V. E. Koopmans \inst{1}\and
        A. R. Offringa \inst{3,1}\and
        M. Mevius \inst{3}\and
        V. N. Pandey \inst{3,1} \\
        S. A. Brackenhoff \inst{1}\and
        E. Ceccotti \inst{1}\and
        B. Ciardi \inst{4}\and
        B. K. Gehlot \inst{1,5}\and
        R. Ghara \inst{6,7}\and
        S. K. Giri \inst{8}\and
        I. T. Iliev \inst{9}\and
        S. Munshi \inst{1}
          \fnmsep
          }

   \institute{Kapteyn Astronomical Institute, University of Groningen, PO Box 800, 9700AV Groningen, The Netherlands\\
              \email{hgan@astro.rug.nl}
         \and
             LERMA (Laboratoire d'Etudes du Rayonnement et de la Mati\`{e}re en Astrophysique et Atmosph\`{e}res), Observatoire de Paris,\\ PSL Research University, CNRS, Sorbonne Universit\'{e}, F-75014 Paris, France
         \and
            The Netherlands Institute for Radio Astronomy (ASTRON), PO Box 2, 7990AA Dwingeloo, The Netherlands
         \and
            Max-Planck Institute for Astrophysics, Karl-Schwarzschild-Stra{\ss}e 1, D-85748 Garching, Germany
         \and
            School of Earth and Space Exploration, Arizona State University, Tempe, Az 85281, USA
         \and
            Astrophysics Research Center (ARCO), Department of Natural Sciences, The Open University of Israel,\\ 1 University Road, PO Box 808, Ra'anana 4353701, Israel
         \and
            Department of Physics, Technion, Haifa 32000, Israel
        \and
            Institute for Computational Science, University of Zurich, Winterthurerstra{\ss}e 190, CH-8057 Zurich, Switzerland
         \and
            Astronomy Centre, Department of Physics and Astronomy, University of Sussex, Pevensey II Building, Brighton BN1 9QH, UK
             }

   \date{Received XXX, 2022; accepted XXX, 2022}


  \abstract
   {Detecting the 21-cm signal from the Epoch of Reionisation (EoR) has been highly challenging due to the strong astrophysical foregrounds, ionospheric effects, radio frequency interference (RFI) and instrumental effects. Better characterisation of their effects and precise calibration are, therefore, crucial for the 21-cm EoR signal detection. 
   }
   {In this work, we introduce a newly developed algorithm \textsc{ddecal} (Direction-Dependent Calibration) and compare its performance with an existing direction-dependent calibration algorithm, \textsc{sagecal}, in the context of the LOFAR-EoR 21-cm power spectrum experiment.}
   {We process one night of data from LOFAR observed by the HBA system. The observing frequency ranges between 114 and 127 MHz, corresponding to the redshift from 11.5 and 10.2. The North Celestial Pole (NCP) and its flanking fields were observed simultaneously in this data set. We analyse the NCP and one of the flanking fields. While the NCP field is calibrated by the standard LOFAR-EoR processing pipeline, using \textsc{sagecal} for the direction-dependent calibration with an extensive sky model and 122 directions, for the RA 18h flanking field, \textsc{ddecal} and \textsc{sagecal} are used with a relatively simple sky model and 22 directions. Additionally, two different strategies are used for the subtraction of very bright and far sources, Cassiopeia A and Cygnus A.}
   {The resulting estimated 21-cm power spectra show that \textsc{ddecal} performs better at subtracting sources in the primary beam region due to the application of a beam model, while \textsc{sagecal} performs better at subtracting Cassiopeia A and Cygnus A. The analysis shows that including a beam model during the direction-dependent calibration process significantly improves its overall performance. The benefit is obvious in the primary beam region. We also compare the 21-cm power spectra results on two different fields. The results show that the RA 18h flanking field produces better upper limits compared to the NCP for this particular observation.}
   {Despite the minor differences between \textsc{ddecal} and \textsc{sagecal} due to the beam application, we find that the two algorithms yield comparable 21-cm power spectra on the LOFAR-EoR data after foreground removal. Hence, the current LOFAR-EoR 21-cm power spectrum limits are not likely to depend on the direction-dependent calibration method. For this particular observation, the RA 18h flanking field seems to produce improved upper limits ($\sim30\%$) compared to the NCP. }


   \maketitle
%
\section{Introduction}
Observation of the 21-cm signal of neutral hydrogen from the Epoch of Reionisation (EoR) is one of the most promising tools to reveal the formation and evolution history of the Universe~\citep{Furlanetto_2006,Morales_2010,Pritchard_2012,Liu_2020}. Many experiments are designed to detect the 21-cm signal from the EoR, including global experiments that aim at measuring the sky-averaged spectrum of the 21-cm signal with a single receiver, such as EDGES\footnote{Experiment to Detect the Global EoR Signature}~\citep{Bowman_2018}, LEDA\footnote{the Large aperture Experiment to detect the Dark Ages, \url{http://www.tauceti.caltech.edu/leda/}}~\citep{Greenhill_2012}, PRIZM\footnote{the Probing Radio Intensity at high Z from Marion}~\citep{Philip_2018} and SARAS\footnote{Shaped
Antenna measurement of the background RAdio Spectrum.}~\citep{Singh_2017,Thekkeppattu_2021}; and interferometric experiments that aim at measuring the spatial brightness-temperature fluctuations of the 21-cm signal with a radio interferometer, such as GMRT\footnote{Giant Metrewave Radio Telescope, http://gmrt.ncra.tifr.res.in}~\citep{Paciga_2011,Paciga_2013}, LOFAR\footnote{Low-Frequency Array, \url{http://www.lofar.org}}~\citep{vanHaarlem_2013,Patil_2017,Mertens_2020}, MWA\footnote{Murchison Widefield Array, \url{http://www.mwatelescope.org}}~\citep{Bowman_2013,Barry_2019b,Li_2019} and PAPER\footnote{the Donald C. Backer Precision Array for Probing the Epoch of Reionisation, \url{http://eor.berkeley.edu}}~\citep{Parsons_2012,Cheng_2018,Kolopanis_2019}, as well as the second generation instruments, HERA\footnote{Hydrogen Epoch of Reionisation Array, \url{http://reionization.org/}}~\citep{DeBoer_2017,HERA_2021} and SKA\footnote{the Square Kilometer Array, \url{http://www.skatelescope.org}}~\citep{Mellema_2013,Koopmans_2015}.

However, the detection of the 21-cm signal has been very challenging, because the observed measurements are contaminated by the astrophysical foregrounds that are about 4-5 orders of magnitude stronger than the expected 21-cm signal~\citep{Bowman_2009,Mertens_2018,Gan_2022}, ionosphere~\citep{Mevius_2016,Vedantham_2016,Edler_2021} and radio frequency interference~\citep[RFI;][]{Offringa_2012,Offringa_2019a}, as well as instrumental effects~\citep{Offringa_2019b}. Hence, suppressing these effects during calibration is crucial for detection~\citep{Barry_2016}.

The calibration of the LOFAR-EoR KSP (Key Science Project) data uses the sky-based calibration approach. The processing pipeline of data has been developed and improved over a decade~\citep{Yatawatta_2013a,Patil_2016,Patil_2017,Mertens_2020,Mevius_2022}. Due to the wide field of view of LOFAR, the data need to be calibrated depending on direction to correct for different errors from the varying beam and ionospheric effects. This direction-dependent (DD) calibration step, in particular, has been carried out by \textsc{sagecal}~\citep{Yatawatta_2011,Yatawatta_2015,Yatawatta_2019}. While \textsc{sagecal} has shown excellent calibration performance, no other DD-calibration code has yet been applied to LOFAR-EoR data.

This study introduces a newly developed DD-calibration algorithm \textsc{ddecal}~\citep[Direction-Dependent Calibration;][]{Diepen_2018} and compares the performance of two DD-calibration algorithms, \textsc{ddecal} and \textsc{sagecal}, in the context of LOFAR-EoR 21-cm power spectra. The two algorithms have some differences, especially in the application of beam and constraining of gain smoothness in frequency. These could result in different calibration performance. To study the differences between the two algorithms, we process one-night of raw data obtained with the LOFAR High-Band Antenna (HBA) system on an unexplored flanking field of the North Celestial Pole (NCP) following similar steps in the standard LOFAR-EoR pipeline~\citep{Patil_2017,Mertens_2020}. We use two different DD-calibration algorithms, \textsc{ddecal} and \textsc{sagecal} with a more limited sky model and fewer directions, compared to the current analysis of the NCP field. The goal of the paper is not to compare the two DD-calibration algorithms using an identical sky model, clustering and settings, but to test the full end-to-end processing in terms of the resulting power spectra when the current-best settings and models for both algorithms are used, within the limits of their implementation. The observation covers the unexplored frequency range, from 114 to 127 MHz corresponding to the redshift range $z=11.5-10.2$, pointing at RA 18$^\text{h}$, DEC +86$^\circ$.

For DD-calibration, we use fewer directions ($\sim20$) compared to the standard $122$ directions used for the NCP analysis~\citep{Patil_2017,Mertens_2020}. The DD-calibration step is performed by two algorithms, \textsc{ddecal} and \textsc{sagecal}.
Besides varying the calibration scheme, we also test a ``peeling'' scheme. The peeling scheme, first proposed by~\cite{noordam2004proc}, calibrates and subtracts bright sources sequentially in decreasing order of brightness.
In~\cite{Gan_2022}, we found that residuals of two very far and bright sources, Cassiopeia A and Cygnus A (Cas~A and Cyg~A, hereafter) may be one of the sources of the excess power in the 21-cm power spectra. In this work, we model and subtract these two bright sources separately from the full sky model to improve the calibration performance. Similar approaches have been taken for the bright sources in~\cite{Patil_2017,Gehlot_2019,Mertens_2020}.

The paper is arranged as follows. In section~\ref{sec:observation}, we describe the data and the observational setup. In section~\ref{sec:dd}, the strategy of the DD-calibration with LOFAR is described in detail and we summarise the two DD-calibration algorithms, \textsc{ddecal} and \textsc{sagecal}. Section~\ref{sec:processing} is dedicated to the description of the processing of LOFAR-EoR data. In section~\ref{sec:results}, we present DD-calibration results with different algorithms and strategies including residual images and power spectra. Different gain smoothness constraints between \textsc{ddecal} and \textsc{sagecal} are discussed in more depth in subsection~\ref{subsub:gain_smoothness}. In section~\ref{sec:conclusions}, we summarise the results and conclude. 
\section{Observation}
\label{sec:observation}
   \begin{table}
      \caption[]{Summary of observational details of L612832.}
         \label{tab:obsv}
     $$
         \begin{array}{p{0.5\linewidth}l}
            \hline
            \noalign{\smallskip}
            Observation ID & \text{L612832} \\
            Observing project & \text{LT5}{\_009} \\
            Pointing (\textit{J2000.0}) & 18^\text{h}00^\text{m}00^\text{s}, \text{+86}^{\circ}00^{\prime}00^{\prime\prime}\\
            Frequency range & 113.8657 \text{-} 127.1469\text{ MHz}\\
            Redshift range & 11.54 \text{-} 10.23\\
            Observation start time (UTC) & \text{2017-10-02  17:33:16.0}\\
            Observation end time (UTC) & \text{2017-10-03  05:11:04.1}\\
            Duration & 41868.1\text{ s } (\sim11.6\text{ h}) \\
            Sub-band width & 183.1 \text{ kHz}\\
            Time, frequency resolution & \\
            \hspace{0.5cm} Before averaging & 2\text{ s},\text{ }3.05\text{ kHz}\\
            \hspace{0.5cm} After averaging & 10\text{ s},\text{ }61.035\text{ kHz}\\
            \noalign{\smallskip}
            \hline
         \end{array}
     $$
   \end{table}

\begin{figure*}
    \centering
    \includegraphics[width=0.6\textwidth]{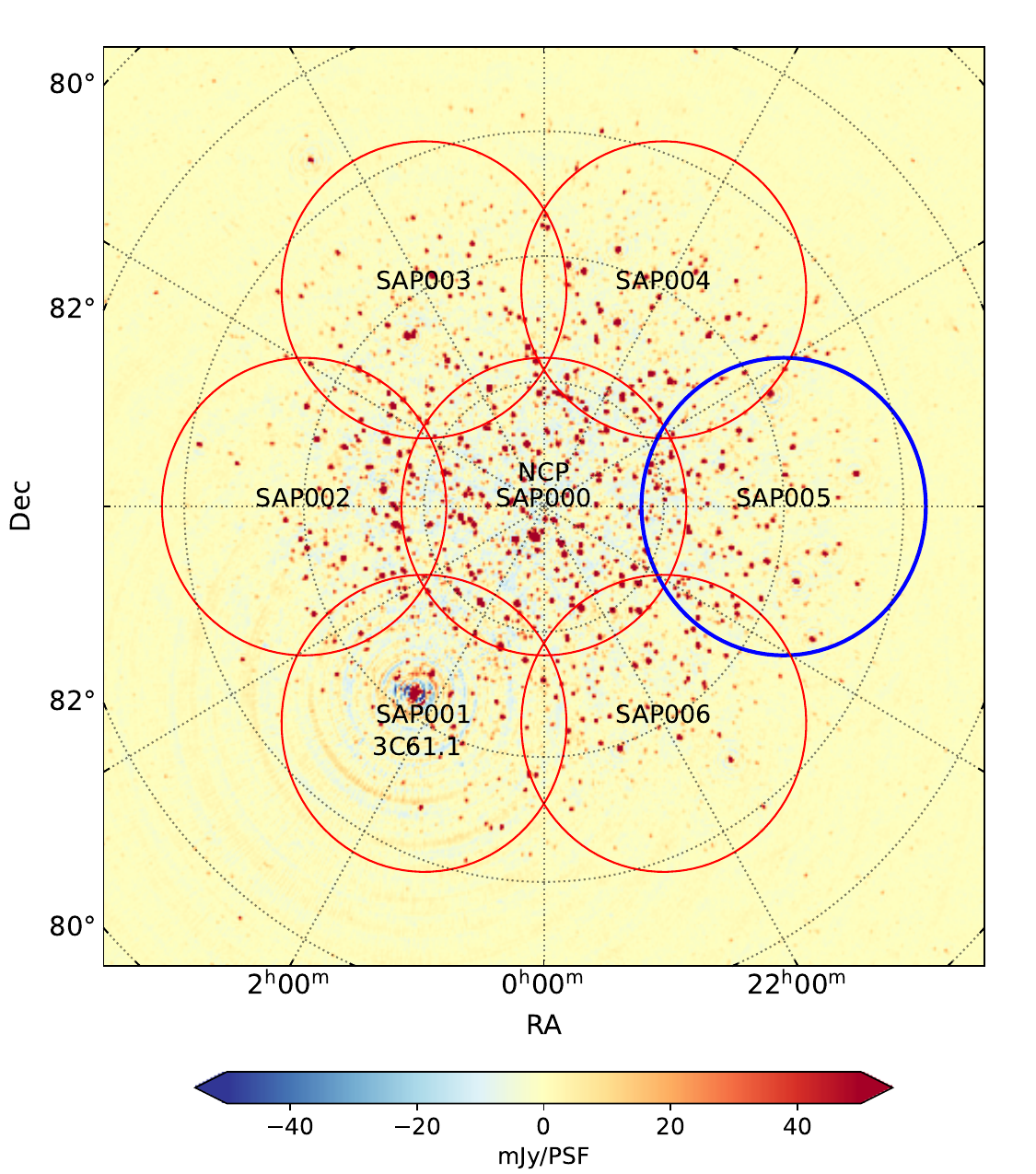}
    \caption{Observing configuration of the NCP field in LOFAR-EoR. The main target field NCP is located in the centre and six flanking fields are distributed from the centre at an angular distance of $4^\circ$ (marked with circles of radius $4^\circ$). The bright radio source 3C61.1 is inside the RA 2h flanking field (SAP001). The image is constructed from a single observation night L612832 ($\sim11.6$-hour long) using full sub-bands. In this work, we analyse one-night data on the RA 18h field (marked with a blue circle).}
    \label{fig:FF_dis}
\end{figure*}
The data analysed in this work have been obtained by the LOFAR High-Band Antenna (HBA) system~\citep{vanHaarlem_2013}. The observational details of the data are summarised in Table~\ref{tab:obsv}. The LOFAR-EoR KSP has two target fields: the NCP~\citep{Yatawatta_2013a} and a field centred on the bright compact radio source 3C196~\citep{Bernardi_2010}. From the LOFAR observation Cycles 0 to 10, about $2450$ hours (more than 100 nights) and $\sim1100$ hours of data have been collected on these two fields, respectively. Around $75\%$ of the collected data are assumed to be of good quality.
In later observation cycles, the main fields have a configuration with a target field in the centre surrounded by a hexagonal ring of six flanking fields. For the NCP field, flanking fields are at an angular distance of $4^\circ$. The observing configuration is shown in Fig.~\ref{fig:FF_dis} with the core station FWHM ($\sim4.8^\circ$ at 120 MHz~\citealp{vanHaarlem_2013}). The pointing directions of the NCP flanking fields are summarised in Table~\ref{tab:ff}. The flanking field data are collected in addition to the main field data to increase the data volume ($\sim6$ times that of the main field per observation) and build a deep and wide sky model for the NCP. In principle, observing six flanking fields enables one to lower the errors in the 21-cm power spectrum by a square root of seven for a fixed bandwidth. The frequency range was chosen based on the best results from~\cite{Patil_2017}. A disadvantage is that by observing multiple fields on a fixed bandwidth, one will limit the range of bandwidth. Because the main field and flanking fields share the same or very similar RFI, ionospheric environment and systematics, the flanking field data can be used for cross-checking the NCP results. The flanking field data are also useful for calibration, e.g. for constructing better sky models for the main field, improving ionospheric modelling and expanding the field of view for polarisation images~\citep{Patil_2017}.

So far, the two published LOFAR-EoR KSP upper limits on the 21-cm signal power spectra in~\citep{Patil_2017,Mertens_2020} are based solely on NCP observations. In this work, for the first time, we analyse one night data on one of the six NCP flanking fields from the LOFAR observation Cycle 5, the RA 18h flanking field (marked with a blue circle in Fig.~\ref{fig:FF_dis}). We create a new sky model on the chosen field, calibrate the data using the built sky model and estimate a 21-cm power spectrum. The 21-cm power spectrum is compared with the one on the NCP field for a cross-check.

While \cite{Patil_2017} and \cite{Mertens_2020} focused on the frequency range from 121.8 to 159.3 MHz (equivalently, z = 10.6 - 7.9), we analyse the frequency range from 113.4 to 127.1 MHz, corresponding to a slightly higher redshift range from 11.5 to 10.2. We chose the RA 18h flanking field for analysis, because this field has never been analysed at this frequency range before. The data are obtained during nighttime to minimise ionospheric effects and avoid the sun, using all core stations and remote stations, with a spectral resolution of 3.05 kHz and a temporal resolution of 2 seconds. The observation duration is around 11.6 hours. The observational details are summarised in Table.\ref{tab:obsv}.

\section{Direction-dependent calibration}
\label{sec:dd}
The propagation of the signal from radio sources to the radio interferometer is often described by the Radio Interferometric Measurement Equation~\citep[RIME,][]{Hamaker_1996,Smirnov_2011A&A...527A.106S}. Considering an array of N elements, the correlation of signals between the $i$-th and $j$-th elements at frequency $\nu$ and time $t$ produces the observed visibility matrix $\textbf{V}_{ij\nu t}$, which can be described as
\begin{equation}
    \textbf{V}_{ij\nu t} = \textbf{J}_{i\nu t}\: \textbf{C}_{ij\nu t} \: \textbf{J}^\text{H}_{j\nu t} + \textbf{N}_{ij\nu t},
\label{eq:RIME}
\end{equation}
where $\textbf{J}_{i\nu t}$ and $\textbf{J}^\text{H}_{j\nu t}$ are $2\times2$ Jones matrices at frequency $\nu$ and time $t$ for element $i$ and $j$. $\textbf{C}_{ij\nu t}$ is a $2\times2$ coherency matrix of the intrinsic signal in a certain direction at the $i$-th and $j$-th elements (i.e., baseline $ij$). The Jones matrices describe the electromagnetic interaction of the intrinsic signal, such as the instrumental effects including the beam shape and receiver response, and propagation effects including ionospheric distortions~\citep{Hamaker_1996,Born_1999}. $\textbf{N}_{ij\nu t}$ is a $2\times2$ noise matrix of baseline $ij$.

Due to the wide field of view of LOFAR\footnote{The LOFAR core station field of view is $\sim17.73\text{ deg}^2$ at 120 MHz~\citep{vanHaarlem_2013}.}, LOFAR data need to be calibrated direction-dependently to compensate for different errors from varying beam and ionospheric effects. The sky model consists of many thousands of bright (a few Jy) and faint (a few mJy) discrete sources. These sources, therefore, need to be clustered to $K$ directions for the direction-dependent (DD) calibration. Each cluster must have a sufficient integrated flux so that a DD gain solution can be obtained in a given time and frequency range with a high enough signal-to-noise ratio. The observed visibility matrix for elements $i$ and $j$ in Eq.~(\ref{eq:RIME}) then replaces the true sky with the sky model, becoming
\begin{equation}
    \textbf{V}_{ij\nu t} = \sum_{k=1}^{K} \textbf{J}_{ik\nu t}\: \textbf{C}_{ijk\nu t} \: \textbf{J}^\text{H}_{jk\nu t} + \textbf{N}_{ij\nu t},
\label{eq:RIME_k}
\end{equation}
where $k$ indicates the specific direction where gains are solved for. The goal of calibration is to estimate a set of parameters $\boldsymbol{\theta}$ describing the Jones matrices at a given time $t$, frequency $\nu$ and element ($i$ or $j$) in Eq.~(\ref{eq:RIME}). The solutions can either be applied to data to correct for the non-signal effects such as the ionosphere and instrumental errors, or the solutions predicted from a sky model can be subtracted from the data to calculate the residuals. Direction-independent (DI) gains are often applied to the data, whereas direction-dependent (DD) gains are used during the subtraction of the sky model.

The parameters $\boldsymbol{\theta}$ can be estimated by minimising the least-square's cost function
\begin{equation}
    \begin{split}
         g(\boldsymbol\theta)
        = \sum_{\nu,t,i,j}
        \Big\| \textbf{V}_{ij\nu t} -
        \sum_{k=1}^{K} \textbf{J}_{ik}(\boldsymbol\theta)\: \textbf{C}_{ijk\nu t} \: \textbf{J}^{\text{H}}_{jk}(\boldsymbol\theta)
        \Big\|^2.
    \end{split}
    \label{eq:cost_fn}
\end{equation}
In the calibration process, the gain solutions are assumed to be invariant over a small but finite time and frequency interval.

One of the main assumptions used for calibration is other effects including instrumental and ionospheric effects are intrinsically smooth as a function of frequency, while the 21-cm EoR signal is not. Enforcing spectral smoothness can, therefore, drastically improve the calibration performance by avoiding overfitting and signal suppression~\citep{Millad_2018,Mevius_2022}. Known spectrally unsmooth effects, such as RFI and cable reflections are handled by (RFI) excision or are treated as a DI bandpass error that can be solved at the DI-calibration step.

There are many calibration algorithms to solve the RIME in Eq.~(\ref{eq:RIME_k})~\citep[e.g.][]{Kazemi_2011,Kazemi_2013,Tasse_2014,OLLIER_2018,Arras_2019}. In this work, we focus on two algorithms, \textsc{ddecal} and \textsc{sagecal}, and compare their performance in the context of LOFAR-EoR 21-cm power spectra.

\subsection{\textsc{ddecal}}
\label{subsec:ddecal}
We use \textsc{ddecal} as one of our DD-calibration tools in the analyses of this work. \textsc{ddecal} is part of the \textsc{dp3}~\citep[Default Preprocessing Pipeline processing software;][]{Diepen_2018}\footnote{The source code for \textsc{dp3} can be found at \url{https://www.astron.nl/citt/DP3/}, and the \textsc{dp3} documentation can be found at \url{https://www.astron.nl/citt/DP3/}}. \textsc{dp3} performs streaming operations on an astronomical data set, such as flagging, averaging, calibration, compression, statistical and various other corrections. \textsc{dp3} is configured by providing a so-called parameter set (\texttt{parset}), which defines the operations to perform, as well as their parameters. \textsc{ddecal} is implemented into \textsc{dp3} with the purpose of having a flexible framework to integrate constrained calibration algorithms. At present, it integrates four algorithms: a directional solving algorithm \citep{smirnov-tasse-2015}; a direction-iterative algorithm \citep{offringa-2016}; the Limited-memory Broyden–Fletcher–Goldfarb–Shanno (LBFGS) algorithm \citep{lbfgs, yatawatta-lbfgs} and a hybrid algorithm that can combine methods. In this work, we have only used the directional-solving algorithm, which we will describe in the next section.

\subsubsection{Directional solving in \textsc{ddecal}}
In each iteration, the directional-solving algorithm finds the solution of all directions for a single element from the measurement equation of Eq.~\eqref{eq:RIME}. It is an extension of the iterative single-directional solve algorithm \citep{rts-mwa-2008,salvini-2014}. If we define $\mathcal{J}_i$ to be a matrix consisting of the $2\times 2$ matrices for element $i$ and all the solved directions, stacked in the column direction:
\begin{equation}
 \mathcal{J}_i = \begin{pmatrix}
     \textbf{J}_{i,k=0} &
     \textbf{J}_{i,k=1} &
     \textbf{J}_{i,k=2} &
     \cdots
 \end{pmatrix},
\end{equation}
the solve algorithm finds the least-squares solution for $\mathcal{J}_i$, i.e., the calibration solutions for a single element but all directions at once:
\begin{equation}
  \mathcal{J}_i
     = \argmin\limits_{\mathcal{J}_i} \sum\limits_{\nu,t,j} \Big\| \textbf{V}_{ij\nu t}\: - \sum\limits_k
    \textbf{J}_{ik\nu t}\: \textbf{C}_{ijk\nu t} \:
    \textbf{J}^\text{H}_{jk\nu t} \Big\|^2.
    \label{eq:ddecal-minimization}
\end{equation}
During one iteration, the solutions for every element are calculated one by one and updated by moving the old value towards the new value, and this is iterated until convergence. This is the algorithm described by \citet{smirnov-tasse-2015}.

To solve Eq.~\eqref{eq:ddecal-minimization}, we define matrices $\mathcal{V}$ and $\mathcal{M}$ that contain the multi-directional data visibilities and corrected model visibilities for one element. When we introduce an index symbol $w$ that enumerates over all values of $\nu$ and $t$ inside the solution interval\footnote{$w$ loops over ($\nu$,$t$) for all possible solution intervals at a given direction $k$ and element $j$. ($\nu$,$t$) alone indicates a solution at a certain interval.}, these two matrices can be defined by

\begin{equation}
\begin{split}
&\mathcal{V}_i = \begin{pmatrix}
    \textbf{V}_{i,j=0,w=0} &
    \textbf{V}_{i,j=0,w=1} &
    \cdots &
    \textbf{V}_{i,j=1,w=0} &
    \cdots &
\end{pmatrix}, \\
&\textbf{M}_{i,j,w,k} =  \textbf{C}_{i,j,w,k}\:\textbf{J}^\text{H}_{jk},\\
&\mathcal{M}_i = \\
&\begin{pmatrix}
\textbf{M}_{i,j=0,w=0,k=0} &
\textbf{M}_{i,j=0,w=1,k=0} &
\cdots &
\textbf{M}_{i,j=1,w=0,k=0} &
\cdots \\
\textbf{M}_{i,j=0,w=0,k=1} &
\textbf{M}_{i,j=0,w=1,k=1} &
\cdots &
\textbf{M}_{i,j=1,w=0,k=1} &
\cdots \\
\cdots
\end{pmatrix},\\
\end{split}
\end{equation}
where the columns of $\mathcal{V}_i$ and $\mathcal{M}_i$ enumerate all combinations of $w$ and $j$ (excluding $i=j$), and the directions are stacked in the rows of $\mathcal{M}_i$. With these definitions, the solution to Eq.~\eqref{eq:ddecal-minimization} is simplified to
\begin{eqnarray} \label{eq:ddecal-step}
\mathcal{V}_{i} &=& \mathcal{J}_{i} \mathcal{M}_{i}.
\end{eqnarray}

This results in a $2 \times 2 N_w N_a$ matrix $\mathcal{V}_i$, a $2 \times 2 N_d$ matrix $\mathcal{J}_i$ and a $2 N_d \times 2 N_w N_a$ matrix $\mathcal{M}_i$, with $N_d$ the number of directions and $N_w$ the number of timesteps $\times$ frequencies inside the solution interval.

Eq.~\eqref{eq:ddecal-step} is a standard linear equation, and $\mathcal{J}_{i}$ can be solved for by standard linear algebra techniques such as using the normal equations $\mathcal{J}_{i} = \mathcal{V}_{i} \mathcal{M}_{i}^H (\mathcal{M}_{i} \mathcal{M}_{i}^H)^{-1}$, or QR-decomposition or singular-value decomposition of $\mathcal{M}_{i}$. \textsc{ddecal} supports these three methods, and we have found that QR-decomposition generally results in a good compromise between accuracy and speed.

Besides the full Jones problem shown here, \textsc{ddecal} has specialisations of this algorithm to find diagonal and scalar solutions, and can optionally constrain the algorithm to find phase-only or amplitude-only solutions, or solve for differential Faraday rotation.

\subsubsection{Applying constraints to the algorithms}
\textsc{ddecal} allows the application of constraints on its four algorithms, including the directional-solving algorithm which is used in this paper. The implemented algorithms are written such that they iteratively step toward the solution. Updated solutions are used in the next iteration, leading again to more accurate solutions (as long as the algorithm converges), which repeats until the accuracy tolerance has been reached. Such iterative algorithms make it relatively easy to find constrained solutions: after moving the solutions towards the direction given by Eq.~\eqref{eq:ddecal-step}, a constraint can be applied.

\textsc{ddecal} allows the application of different types of constraints, including spatial, temporal and spectral constraints. In this work, we use a constraint that forces the solutions to be spectrally smooth. \textsc{ddecal} implements this by Gaussian smoothing the solutions with a requested width. When applying a spectral smoothness constraint, \textsc{ddecal} calculates the next solution step independently for a number of channels, applies the smoothness constraint to all solutions simultaneously and then continues with the next iteration for each channel, repeating this until the channels simultaneously reach the stopping criterion.

\subsection{\textsc{sagecal}}
\label{subsec:sagecal}
The space alternating generalised expectation maximisation (SAGE) algorithm~\citep{Fessler_1994,Kazemi_2011} can be used to estimate the parameters describing $\textbf{J}_{ik}$ for all possible values of $i$ and $k$ in Eq.~\ref{eq:RIME_k}.

\subsubsection{SAGE algorithm}
The `expectation' step of the SAGE algorithm calculates the effective observed data along the $m$-th direction in a finite time interval, using
\begin{equation}
    \textbf{V}_{ijm\nu} =
    \textbf{V}_{ij\nu} - \sum_{k=1, k\neq m}^{K}
    \hat{\textbf{J}}_{ik\nu}\: \textbf{C}_{ijk\nu}\:
    \hat{\textbf{J}}^\text{H}_{jk\nu},
\end{equation}
where $\hat{\textbf{J}}_{ik\nu}$ and $\hat{\textbf{J}}^\text{H}_{jk\nu}$ are the estimated Jones matrices. The `maximisation' step minimises the objective function only for the $m$-th direction defined under a Gaussian noise model as
\begin{equation}
\begin{split}
     g_{m\nu} \left(\textbf{J}_{1m\nu}, \textbf{J}_{2m\nu}, ...  \right)
     = \sum_{i,j} \Big\| \textbf{V}_{ijm\nu}\: -
    \textbf{J}_{im\nu}\: \textbf{C}_{ijm\nu} \:
    \textbf{J}^\text{H}_{jm\nu} \Big\|^2.
    \label{eq:maximisation}
\end{split}
\end{equation}
Using the SAGE algorithm, Eq.~\ref{eq:RIME_k} can be simplified from a simultaneous calibration along $K$ directions to $K$ single direction sub-problems~\citep{Kazemi_2011,Yatawatta_2016fine}. For simplicity, below, we consider the calibration along one direction only and drop the subscript $m$, such that Eq.~\ref{eq:maximisation} becomes
\begin{equation}
\begin{split}
      g_\nu \left( \textbf{J}_\nu  \right)
      = \sum_{i,j} \Big\| \textbf{V}_{ij\nu}\: -
    \textbf{A}_i \textbf{J}_\nu\: \textbf{C}_{ij\nu} \:
    \left( \textbf{A}_j \textbf{J}_\nu  \right)^\text{H} \Big\|^2,
    \label{eq:objective}
\end{split}
\end{equation}
where $\textbf{J}_\nu$ is Jones matrices for all elements along the $m$-th direction and $\textbf{A}_i$ is the canonical selection matrix to choose $i$-th element among N elements,
\begin{equation}
\begin{split}
    & \textbf{J}_\nu \triangleq
    \left[ \textbf{J}_{1m\nu}^\text{T},
    \textbf{J}_{2m\nu}^\text{T}, ...,
    \textbf{J}_{Nm\nu}^\text{T}
    \right]^\text{T}, \\
    & \textbf{A}_i \triangleq
    \left[ \textbf{0}, \textbf{0}, ..., \mathrm{I}, ..., \textbf{0}\right],
    \label{eq:J_A_def}
\end{split}
\end{equation}
where only the $i$-th matrix of $\textbf{A}_i$ is an identity matrix divided in time or frequency in Eq.~\ref{eq:J_A_def}.

\subsubsection{Applying constraints to solutions}
An observation with $P$ data sets is distributed over $C$ compute agents (typically, $P \gg C$). Each data set has several frequency channels and each channel can be identified by its central frequency. Given that all known effects are spectrally smooth, \textsc{sagecal} constrains the continuity of $J_\nu$ over frequency to improve the calibration performance by applying the consensus alternating direction method of multipliers algorithm \citep[C-ADMM;][]{Boyd_2011,Yatawatta_2015,Yatawatta_2016fine}. 
The objective function in Eq.~\ref{eq:objective} is then modified to an augmented Lagrangian with a regularisation parameter to guide solutions to approach the smooth regularisation function of choice $\textbf{B}_\nu \textbf{Z}$,
\begin{equation}
   \mathcal{L}_\nu \left( \textbf{J}_\nu, \textbf{Z},
   \textbf{Y}_\nu \right) =
   g_\nu \left( \textbf{J}_\nu  \right) +
   \| \textbf{Y}^\text{H}_\nu (\textbf{J}_\nu -
   \textbf{B}_\nu \textbf{Z}) \|
   +\frac{\rho}{2} \| \textbf{J}_\nu - \textbf{B}_\nu \textbf{Z}  \|^2,
   \label{eq:admm}
\end{equation}
where $\mathcal{L}_\nu$ denotes the Lagrange multiplier and the continuity of frequency is constrained by the frequency model described by a set of basis functions $\textbf{B}_\nu$. \textsc{sagecal} uses third order Bernstein polynomials~\citep{Farouki_1988} as the basis functions~\citep{Yatawatta_2019}. $\textbf{Z}$ is a global variable shared by all frequencies in the data.
The $n$-th ADMM iteration solves Eq.~\ref{eq:admm} in the following three steps for all frequencies in parallel:
\begin{equation}
    ( \textbf{J}_\nu )^{n+1} =
    \arg \min_\textbf{J} \mathcal{L}_\nu
    \Big( \textbf{J}, (\textbf{Z})^n, (\textbf{Y}_\nu )^n \Big),
\end{equation}
\begin{equation}
    ( \textbf{Z} )^{n+1} =
    \arg \min_\textbf{Z} \sum_{\nu}
    \mathcal{L}_\nu
    \Big( (\textbf{J}_\nu)^{n+1}, (\textbf{Z}), (\textbf{Y}_\nu )^n \Big),
\end{equation}
\begin{equation}
    ( \textbf{Y}_\nu )^{n+1} =
    (\textbf{Y}_\nu)^n  +
    \rho \Big( (\textbf{J}_\nu)^{n+1} - \textbf{B}_\nu (\textbf{Z})^{n+1} \Big),
\end{equation}
where the superscript $(.)^n$ denotes the $n$-th iteration and $\rho$ is a regularisation parameter that determines the level of smoothness in frequency for each iteration. For our observation, $\rho\sim1000$ is found to be optimal for 30 ADMM iterations. For more discussions about the selection of $\rho$, we refer readers to \cite{Yatawatta_2015,Yatawatta_2016fine,Mertens_2020,Mevius_2022}.

\subsection{Differences between \textsc{ddecal} and \textsc{sagecal}}
Mathematically, both \textsc{ddecal} and \textsc{sagecal} find gain solutions by minimising the least square's cost function given by Eq.~\ref{eq:cost_fn}. Their detailed implementations, however, are different.

In this work, \textsc{ddecal} applies the direction-solving algorithm to find solutions for all directions at a given element (i.e. an antenna), while \textsc{sagecal} applies the SAGE algorithm to find solutions for all elements at a fixed direction.

The frequency smoothness of gains is constrained differently in the two methods. \textsc{ddecal} smooths gains by convolving them with a Gaussian kernel of a chosen bandwidth during each iteration of the optimisation, while \textsc{sagecal} uses an augmented Lagrangian with a regularisation parameter to enforce the gain smoothness.

Another important difference between \textsc{ddecal} and \textsc{sagecal} is the application of beam. \textsc{ddecal} supports the LOFAR HBA station beam (with \texttt{usebeammodel} in \textsc{dp3}), for that reason, an intrinsic sky model is used for calibration. \textsc{sagecal} currently does not support the LOFAR station beam model\footnote{The latest version of \textsc{sagecal} only supports the LOFAR dipole beam model, in the future, more beam options will be supported; \url{http://sagecal.sourceforge.net/}}. Hence, an apparent sky model (which folds the average beam into the sky model) is used for calibration.

In theory, DD-calibration is supposed to solve gains for an optimal number of directions and solution intervals to take care of beam variations and ionospheric phase shifts. However, this process is not perfect and there are errors. In this work, \textsc{ddecal} uses an intrinsic sky model with an HBA beam model, while \textsc{sagecal} uses an apparent model without a beam model.
We focus on how these differences affect the results of DD-calibration and 21-cm signal power spectra.

\section{Outline of the data processing}
\label{sec:processing}
\begin{figure*}
    \centering
    \includegraphics[width=\textwidth]{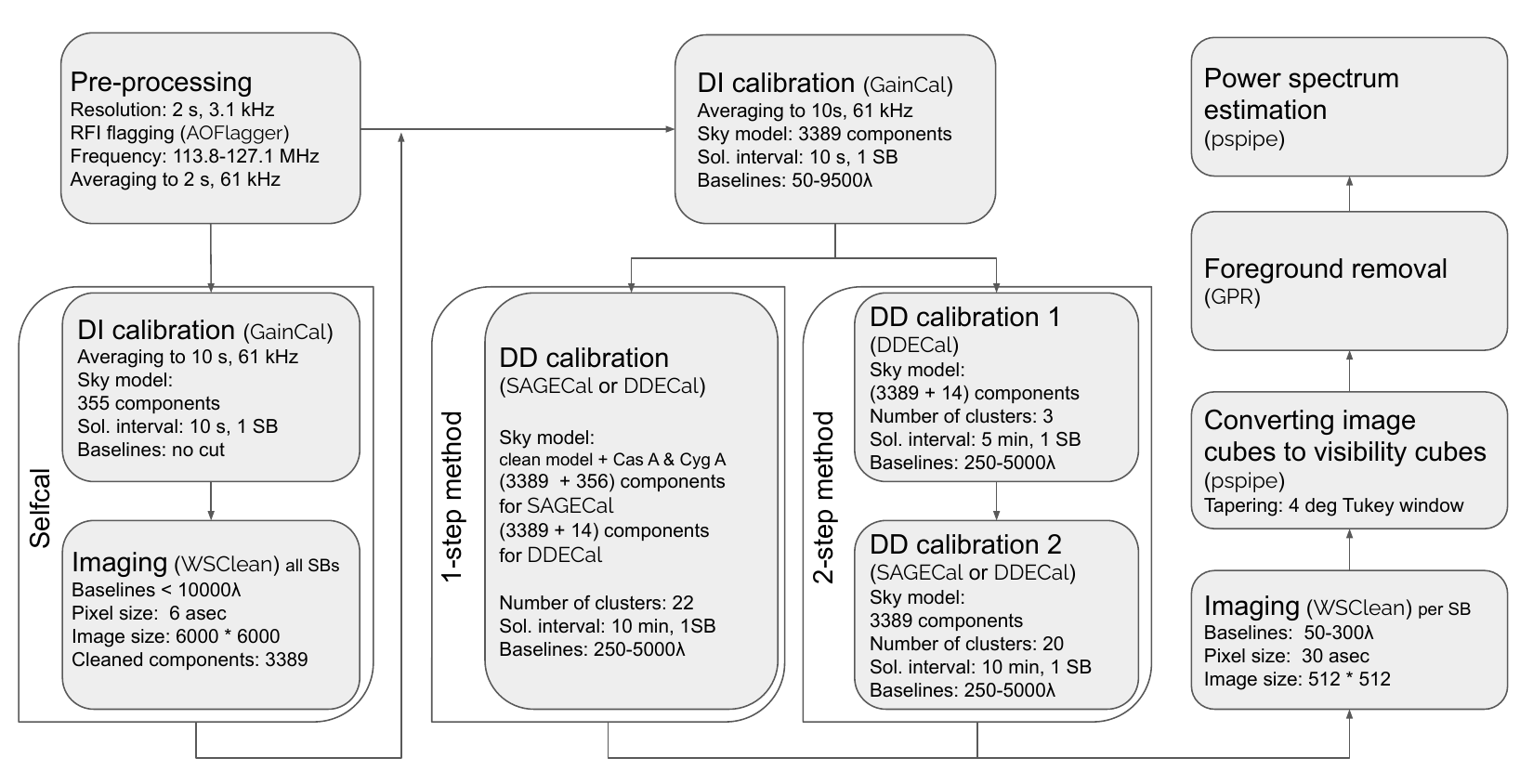}
    \caption{Processing pipeline of the LOFAR-EoR flanking field data in this work to obtain the 21-cm power spectra with two different DD-calibration algorithms. The pipeline is slightly different from the standard LOFAR-EoR HBA processing pipeline in~\citep{pandey_2020,Mertens_2020} because our main purpose is comparing the performance of two DD-calibration algorithms. We also adopt two different clustering approaches for the DD-calibration, i.e. the 1-step and 2-step methods, to investigate how different clustering impacts the subtraction of the sky model.}
    \label{fig:processing}
\end{figure*}
The observed data are processed on the dedicated high performance computing (HPC) cluster \textsc{dawn} which consists of 124 NVIDIA K40 GPUs~\citep{Patil_2017,pandey_2020}. Because we are analysing data on a flanking field of the NCP and our main purpose is to compare the performance of two different DD-calibration algorithms, we use a slightly different data processing strategy from the LOFAR-EoR data processing pipeline adopted in~\cite{Patil_2017,Mertens_2020}, especially for the DI and DD-calibration steps. The main processing steps in this work are: (1) Pre-processing including data averaging and RFI flagging, (2) self-calibration iterations and imaging to create a sky model, (3) averaging and DI-calibration to correct the flux of sources, (4) DD-calibration with the two different algorithms described in Subsections~\ref{subsec:ddecal} and~\ref{subsec:sagecal} to subtract the sky model, (5) imaging, (6) visibility cube conversion, (7) residual foreground removal and (8) power spectrum estimation. Fig.~\ref{fig:processing} shows an overview of the RA 18h flanking field data processing pipeline of this work. Each step will be described in more detail in the following.

\subsection{Pre-processing}
RFI-flagging is performed by \textsc{aoflagger}~\citep{Offringa_2012} on the highest time and spectral resolution of 2 seconds and 3.05~kHz (64 channels per sub-band, 183 kHz). In this step, the four edge channels 0, 1, 62 and 63 of sub-bands are flagged to avoid aliasing effects from the poly-phase filter~\citep{Patil_2017}. After the first RFI-flagging, the remaining 60 channels are averaged to 15 channels (12.2~kHz per channel) and the data are archived in the LOFAR LTA at SURFsara in Amsterdam and Poznan in Poland~\citep{Mertens_2020}. On this averaged data, we perform a second RFI flagging and subsequently the data are averaged to three channels (61 kHz per channel) per sub-band and 2 seconds resolution. This initial RFI-flagging results in a $\sim$5\% loss of the LOFAR-EoR HBA data~\citep{Offringa_2013}. This pre-processing step is identical to the one in the standard LOFAR-EoR pipeline~\citep{Mertens_2020}.

\subsection{Direction-independent calibration}
The first step of calibration begins with a self-calibration~\citep{Cornwell_1981MNRAS.196.1067C,Pearson_1984} and the main goal is to correct the source fluxes and build up a sky model for the DD-calibration. We perform a first gain calibration on the averaged visibilities with a sky model consisting of 355 bright point sources from the NCP sky model.
The gain calibration is carried out by \textsc{gaincal} in \textsc{dp3}. Using \texttt{usebeammodel} option in \textsc{gaincal}, we apply the LOFAR-HBA beam model during calibration and use the initial sky model with intrinsic fluxes. To reduce the data volume and accelerate the calibration process, we first average the data to a 10-second time resolution and gain solutions are calculated on the same time scale per sub-band with the LOFAR-HBA beam model. \footnote{In the standard LOFAR-EoR pipeline, the DI-calibration is conducted on the high resolution data, before averaging. We also performed a test on calibrating the higher resolution data before averaging, but the results were almost identical to the one after averaging. In this case, calibrating on the higher resolution increases the computing time by a factor of 4-5 without a significant improvement. Hence, in this work, we decide to calibrate the data after averaging.} Note that baselines are not limited during the first DI-calibration of self-calibration. Based on our test, if we use the same 50$\lambda$ cut both for self-calibration and the subsequent DI-calibration, data on baselines close to the 50$\lambda$ cut are not well calibrated. For this reason, we decided not to apply a baseline cut when creating a sky model (during self-calibration). The 50$\lambda$ cut is applied during DI-calibration after the self-calibration step. We combine all 69 sub-bands and limit baselines up to 10000$\lambda$ to create a high resolution image with a pixel size of 6 \text{arcsec}. The calibrated visibilities are imaged and deconvolved by the multiscale \texttt{CLEAN} feature of \textsc{wsclean}~\citep{Offringa_2014,Offringa_2017}. 
The obtained \texttt{CLEAN} components are saved as two types of sky models: an apparent model and an intrinsic model. The sources in the apparent model are attenuated by the average beam. The two models will be used in the next calibration steps with combinations of two DD-calibration algorithms. \textsc{ddecal} can apply the LOFAR-HBA beam model in calibration and so we can use an intrinsic sky model. \textsc{sagecal} requires an apparent sky model, because the beam model is not applied.

At this stage, we compare the intrinsic flux of four known bright sources (J190401.7+8536, 6C B184741+851139, 6C B174711+844656 and 6C B163113+855559) around the phase centre (ideally, within $\sim4.75^\circ$, being the FWHM of the LOFAR core stations at 120 MHz;~\citealp{vanHaarlem_2013}) to the ones from catalogues to check whether their fluxes match. The details of sources used for flux scaling and their catalogues are summarised in Table.\ref{tab:flux_scaling}. We aim for the intrinsic flux calibration accuracy of $10\%$ or better. An additional calibration factor is applied to match the intrinsic sky model to the catalogue fluxes. Finally, we perform a DI-calibration using the extended \texttt{CLEAN} component model on the pre-processed data. It is similar to the first DI-calibration step. The data is first averaged to a 10-second time resolution and the \texttt{CLEAN} component model (intrinsic) is used with the LOFAR-HBA beam model. Baselines are limited to $50$-$9500\lambda$ this time. The lower baseline cut is applied to avoid the diffuse emission~\citep{Patil_2017}, while the upper baseline cut comes from the constraint (of 10000$\lambda$) of the sky model.

\subsection{Direction-dependent calibration}
After we have created the new sky model and re-scaled the flux in the DI-calibration step, we perform a DD-calibration. The main goal is to subtract sources in the sky with their DD-calibration gains. In this work, we perform this task with two different DD-calibration algorithms, \textsc{ddecal} and \textsc{sagecal}, described in Subsections~\ref{subsec:ddecal} and \ref{subsec:sagecal}. Because Cas~A and Cyg~A are very bright and far away from the phase centre and their solutions can be distinctive from ones of the remaining sources close to the phase centre. Hence, they need to be calibrated in a separate cluster, similar to the approach used for calibrating the bright source 3C61.1 in the NCP field by~\citet{Patil_2017} and~\citet{Mertens_2020}. Hence, we test two methods for the subtraction of the sky model: (1) the 1-step method: the \texttt{CLEAN} component model with a Cas~A and Cyg~A model are divided into 22 clusters (20 clusters for the \texttt{CLEAN} model and one cluster each for Cas~A and Cyg~A), and all the sources are predicted and subtracted simultaneously in one step; (2) the 2-step method: the \texttt{CLEAN} model, Cas~A and Cyg~A are divided into 3 clusters, respectively. Cas~A and Cyg~A are predicted and subtracted first, after which the \texttt{CLEAN} model is again divided into 20 clusters, predicted and subtracted from the data. Solutions are calculated for 10-min time intervals and each sub-band for the two DD-calibration algorithms and two different approaches to subtracting the sources.

We adopt the same baseline cut applied in the standard LOFAR-EoR pipeline, i.e. 250-5000$\lambda$. The lower baseline cut is used to reduce signal suppression on the baselines of 50-250$\lambda$ used for the 21-cm signal power spectrum extraction, avoid the effects from the diffuse emission and include enough baselines for the required S/N ratio. The upper baseline cut is applied to avoid sky model error and ionospheric phase fluctuations on longer baselines leaking into the short baseline gain solutions~\citep{Patil_2016,Mertens_2020,Mevius_2022}.

\subsection{Imaging and conversion to brightness temperature}
After DI and DD-calibration, imaging, removal of residual foreground and power spectrum estimation are similar to the standard LOFAR-EoR pipeline~\citep[see][for more details]{Mertens_2020}. The residual visibilities after the DD-calibration are gridded and imaged per sub-band to create an image cube using \textsc{wsclean}. We adopt identical imaging parameters used by~\cite{Mertens_2020}, a Kaiser-Bessel anti-aliasing filter with a kernel size of 15 pixels, an oversampling of 4096 and 32 w-layers. According to~\cite{Offringa_2019b}, these parameters are chosen to confine the systematics from gridding below the predicted 21-cm signal.

At this stage, we create even and odd 10-second time-differenced images to estimate the thermal noise of the data. We estimate the thermal noise for the NCP and RA 18h flanking field, and use them for the flux scale cross-check.
The power spectrum is corrected by a factor of two downwards to account for the increase in noise level due to the differencing. The results are discussed in more detail in the following section.

The image cube with a field of view of $12^\circ\times12^\circ$ and a pixel size of 0.5 arcmin is then multiplied by a Tukey function with a diameter of $4^\circ$ to concentrate on the beam centre. The image cube has units of Jy/PSF. For estimating the power spectrum, the image cube is spatially Fourier transformed into a gridded visibility cube and converted to units of Kelvin, as described in~\cite{Offringa_2019b}.

\subsection{Foreground removal and power spectrum estimation}
The remaining foregrounds in the residual Stokes-I visibilities are further removed by the Gaussian Process Regression (GPR) foreground removal technique~\citep{Mertens_2018,Mertens_2020}. GPR enables a separation between different components in observations including smooth astrophysical foregrounds, mode-mixing contaminants, noise and the 21-cm signal by modelling each of them as a Gaussian Process (GP), assuming they can be described by Gaussian processes to first order. GPR properly accounts for degeneracies between the signal components, by marginalising other components.

Finally, the variations of the 21-cm signal as a function of wavenumber $k$ (at different scales) are obtained by a power spectrum. It is estimated by taking the Fourier transform of the foreground-subtracted visibility cube in the frequency direction and converting angle and frequency, to comoving distances~\citep{Morales_2004,McQuinn_2006}.

We can average the power spectrum in $k$-bins to create the spherically-averaged dimensionless power spectrum or define the cylindrically-averaged power spectrum, as a function of angular $k_\perp$ versus line-of-sight $k_{||}$.
\section{Results}
\label{sec:results}
In this section, we present the results of processing one night observation from LOFAR-EoR with \textsc{ddecal} and \textsc{sagecal}. We present the results of each processing step, following the data processing pipeline introduced in section~\ref{sec:processing}. We also compare differences in their performance in terms of removing sources in subsections~\ref{subsec:DD_cal}-\ref{subsec:PS}. In section~\ref{subsec:NCP}, we compare sky images, power spectra and upper limits on the RA 18h flanking field and NCP field.

\subsection{The RA 18h flanking field sky model and DI-calibration}

\begin{table}
      \caption[]{The parameter setup for the multiscale deconvolution algorithm with \textsc{wsclean}.}
         \label{tab:wsclea_pars}
    \centering
\begin{tabular}{l | c}     
\hline
 Parameter & Value \\
\hline
Pixel scale & 6 \text{arcsec}\\
Briggs weighting & $0.0$\\
Baselines & $<10000\lambda$\\
Fitting spectra$^\text{*}$ & 3 terms\\
Auto mask & $7\sigma$\\
Final threshold & $3\sigma$\\
\hline
\end{tabular}
\tablefoot{$^\text{*}$\textsc{wsclean} has an option to enforce a smooth spectrum during joined channel deconvolution by fitting a polynomial. In this work, we fit a polynomial with 3 terms (i.e., a second-order polynomial) to achieve a smooth spectrum.}
\end{table}

The sky model of the RA 18h flanking field is built by the multiscale deconvolution algorithm of \textsc{wsclean} with the parameters listed in Table.\ref{tab:wsclea_pars}.
The intrinsic flux of the \texttt{CLEAN} model is then scaled to match the fluxes of the four bright sources around the phase centre listed in Table.\ref{tab:flux_scaling} at the central observing frequency 119.725~MHz with a spectral index $\alpha=-0.6$. With a flux scaling factor 1.91, we find a mean ratio of 0.997 between the intrinsic \texttt{CLEAN} flux and the references with a standard deviation of 0.0861. In addition, we compare the estimated thermal noise on the RA 18h flanking field and NCP (scaled by NVSS~J011732+892848~\citealp{Patil_2017,Mertens_2020}) from the same observation.
Their estimated thermal noise should closely match if the absolute flux scale is performed accurately. The average ratio of the estimated thermal noise between the NCP and RA 18h flanking field is found to be 1.06, showing that the absolute flux scaling is well performed and the estimated thermal noise values on the two fields are comparable. Note that due to the 4$^\circ$ difference in pointing, the sensitivity is slightly different between the two fields. The noise in part is set by the total power in the beam, which is also partly contributed by sources in the field and diffuse emission. Hence, a perfect flux agreement is not expected. The top panel of Fig.\ref{fig:DD_images} shows images of the RA 18h flanking field after DI-calibration.

\subsection{Direction-dependent gain calibration}
\label{subsec:DD_cal}
In the DD-gain calibration step, we cluster the sky model into a number of directions, predict visibilities in each direction and subtract the clustered sky model sources with their DD-gain applied from the data. An example of the obtained DD-gain power spectra for one station is presented in Appendix.\ref{app:gains}. In this subsection, we will discuss the sky model we use for the DD-calibration and compare its performance using two algorithms, \textsc{ddecal} and \textsc{sagecal}, and two different approaches regarding the subtraction of Cas~A and Cyg~A.

\subsubsection{Clustering of the sky model}
\begin{figure*}
    \centering
    \includegraphics[width=330px]{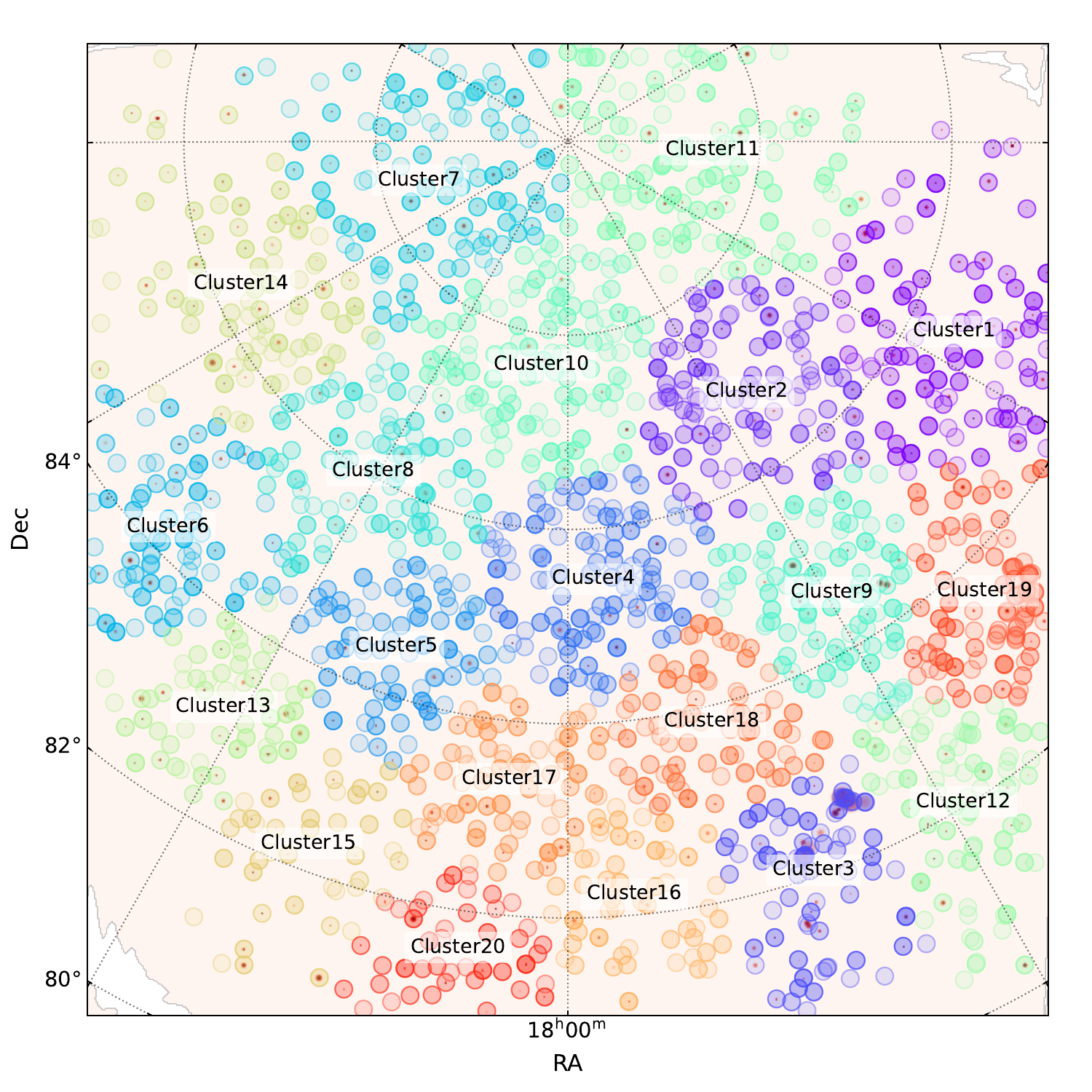}
    \caption{The flanking field sky model from \texttt{CLEAN} components with the LOFAR beam applied clustered into 20 directions for the DD-calibration. Different colours denote different solving directions. Each cluster has an angular radius of $1-2^\circ$.}
    \label{fig:clutering}
\end{figure*}

The 3389 \texttt{CLEAN} components after self-calibration are clustered into 20 directions identically for the two algorithms as shown in Fig.~\ref{fig:clutering}. We make sure that clustering does not contribute to the DD-calibration difference between the two algorithms. The detailed clustering information of the sky model is summarised in Table.\ref{tab:model} and the differences between the two sky models are summarised in Table.\ref{tab:two_models}. Finally, we add Cas~A and Cyg~A into the two sky models, as these bright radio sources, even located outside the field of view, will enter via side-lobes and leave residuals in the power spectrum if not included in the sky model~\citep{Patil_2017,Mertens_2020}. For \textsc{ddecal}, we use the Cas~A and Cyg~A model ($14$ components) from the low resolution A-team sky model \footnote{\url{https://github.com/lofar-astron/prefactor/tree/master/skymodels}}. For \textsc{sagecal}, we use shapelet models created from wide-band LOFAR-LBA and HBA observations with $\sim350$ components~\citep{Yatawatta_2011}\footnote{We also tested the calibration performance with a high resolution Cas~A and Cyg~A sky model with more components with \textsc{ddecal}. However, using more components did not significantly improve the subtraction of the sources. Hence, we decide to use the low resolution model to reduce the computing time.}. The additional Cas~A and Cyg~A components are clustered into their respective directions.

\begin{table*}
      \caption[]{The sky model setups for \textsc{ddecal} and \textsc{sagecal}.}
         \label{tab:two_models}
    \centering
\begin{tabular}{r | c | c }     
\hline
 Parameter & \textsc{ddecal} &  \textsc{sagecal}\\
\hline    \hline
\texttt{CLEAN} model flux & intrinsic & apparent \\\hline
Beam & applied & not applied \\  \hline
Frequency smearing correction & not applied & applied \\ \hline
Time smearing correction & not applied & applied \\ \hline \hline
Number of clusters & \multicolumn{2}{c}{20}  \\ \hline
Number of components & \multicolumn{2}{c}{3389}  \\ \hline \hline
Cas~A \& Cyg~A & Gaussian \& point & shapelet\\
model & sources & sources \\ \hline
Number of clusters & \multicolumn{2}{c}{2}  \\ \hline
Number of components & $14$ & $\sim350$ \\
\hline
\end{tabular}
\end{table*}

\subsubsection{Images}
\label{subsec:images}

\begin{figure*}
   \centering
   \includegraphics[trim={0 0 0 0},clip,width=0.49\textwidth]{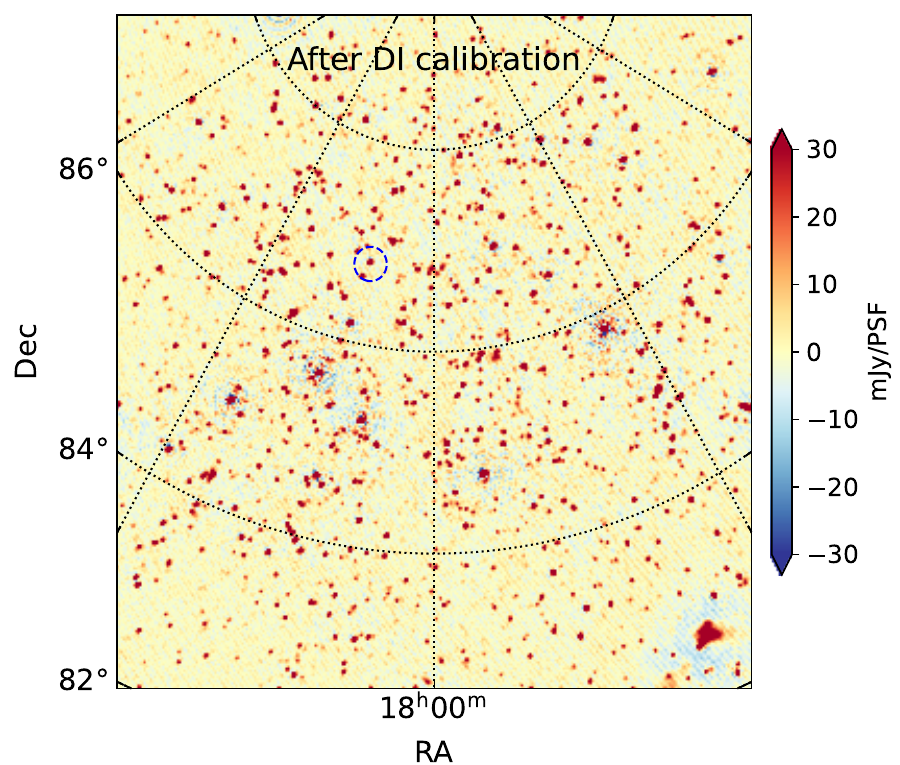}
   \includegraphics[trim={0 0 0 25},clip,width=0.9\textwidth]{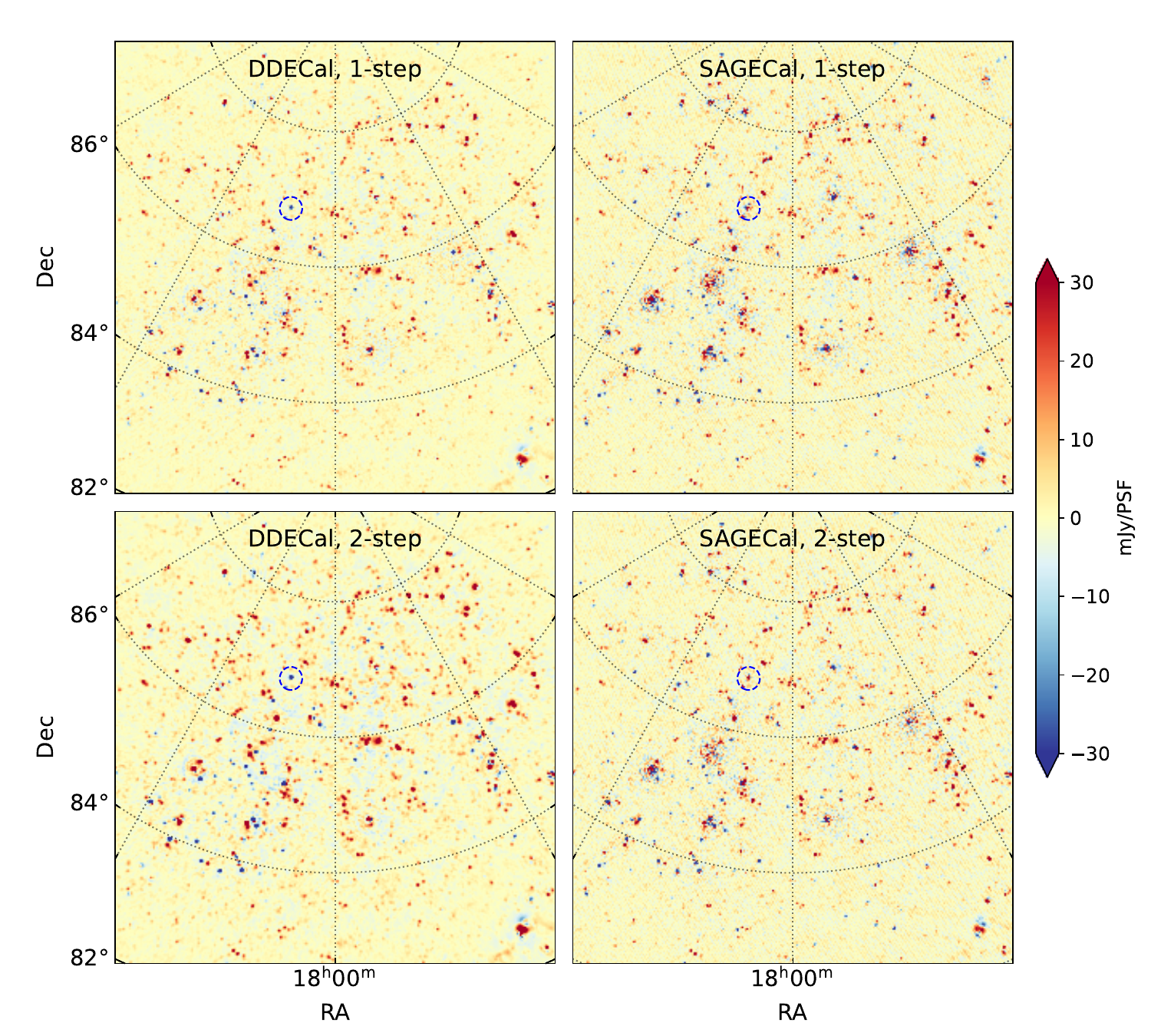}\\
      \caption{LOFAR-HBA $5^\circ\times5^\circ$ Stokes-I residual images after DI- and DD-calibration with four different calibration scenarios on the RA 18h flanking field at frequency 113.9-127.1~MHz. The images are created with a pixel size of 0.2~arcmin using baselines $50-5000\lambda$, combining 69 sub-bands and a single observation night L612832 ($\sim11.6$-hour). \textbf{Top:} after DI-calibration. \textbf{Middle:} calibrated by \textsc{ddecal} (left) and \textsc{sagecal} (right) with the 1-step method). \textbf{Bottom:} calibrated by \textsc{ddecal} (left) and \textsc{sagecal} (right) with the 2-step method). Different DD-calibration scenarios also show different residuals. A source close to the centre is marked with a dashed blue circle as a reference. The residuals of the reference source look different in the four scenarios.
              }
         \label{fig:DD_images}
   \end{figure*}

In Fig.~\ref{fig:DD_images}, we show $5^\circ\times5^\circ$ images of the Stokes-I residuals after DD-calibration with four different scenarios using \textsc{ddecal} and \textsc{sagecal} (middle and bottom).
Compared to the Stokes-I images before DD-calibration in Fig.~\ref{fig:DD_images} (top), most bright sources are removed well after DD-calibration in Fig.~\ref{fig:DD_images} (middle and bottom). In the primary beam region, \textsc{ddecal} (middle left and bottom left in Fig.~\ref{fig:DD_images}) removes more power compared to \textsc{sagecal} (middle right and bottom right in Fig.~\ref{fig:DD_images}) for both 1-step and 2-step methods.

However, depending on the strategy, there are some differences in their residuals. In Fig.~\ref{fig:DD_images}, the images calibrated by \textsc{ddecal} (middle left and bottom left) have more compact residual sources than the images calibrated by \textsc{sagecal} (middle right and bottom right). Notably, \textsc{ddecal} shows better performance with the 1-step method in the primary beam region and the residual power is lower with the 1-step method (middle left) than with the 2-step method (bottom left). The difference between the 1-step and 2-step methods is marginal for \textsc{sagecal} (middle right and bottom right).

We choose a reference source close to the centre to compare the residuals after DD-calibration with the four scenarios. The reference source is marked with a dashed blue circle in Fig.~\ref{fig:DD_images}. The flux of the source is largely reduced after DD-calibration in all four scenarios. \textsc{ddecal} shows an over-subtraction where the source appears as negative (in blue). The over-subtraction is stronger in the 2-step method than in the 1-step method. On the other hand, the residuals of \textsc{sagecal} are rather positive (in red) and not as compact as ones from \textsc{ddecal}. Difference images between the DD-calibration scenarios subtracted by the \textsc{ddecal} and 1-step scenario (middle left) are shown in Fig.\ref{fig:diff_DD}.

Overall, \textsc{ddecal} shows better performance than \textsc{sagecal} in the primary beam region, especially when carried out by the 1-step method. This difference between \textsc{ddecal} and \textsc{sagecal} may be explained by the application of the LOFAR-HBA beam in \textsc{ddecal} which enables more realistic prediction and subtraction of visibilities.

In principle, the number of directions and the time or frequency interval of solutions in DD-calibration are chosen to naturally capture the direction-dependent effects including beam variations and ionospheric phase fluctuations etc. However, this calibration process is not perfect (e.g., due to the incomplete sky model or bad choice of solution intervals) and there are errors. Increasing the solution resolution, i.e., using a finer time or frequency interval for solutions, can be useful for capturing rapid varying beam variations and ionospheric fluctuations to a certain extent, however, it could also introduce extra noise into data and add extra structures in time and frequency. This point is also partially shown in Fig.~\ref{fig:cc_gain_spectra} where we use different time intervals, 5 min and 10 min, to calibrate Cas~A and Cyg~A. The 5 min interval results (bottom panel) did not show an improvement compared to the 10 min interval results (second panel). What we found in this work is that having a physical beam model during DD-calibration will improve the performance of calibration, especially, in the primary beam region.

To compare the subtraction performance of distant sources, such as Cas~A and Cyg~A, we create full sky Stokes-I residual images ($120^\circ\times120^\circ$) after DI-calibration and four different DD-calibration scenarios. The images are created by combining all sub-bands, integrating the full observation and applying $50-300\lambda$ baseline cut (comparable to the cut used for the power spectrum estimation later).

\begin{figure*}
    \centering
    \includegraphics[width=\textwidth]{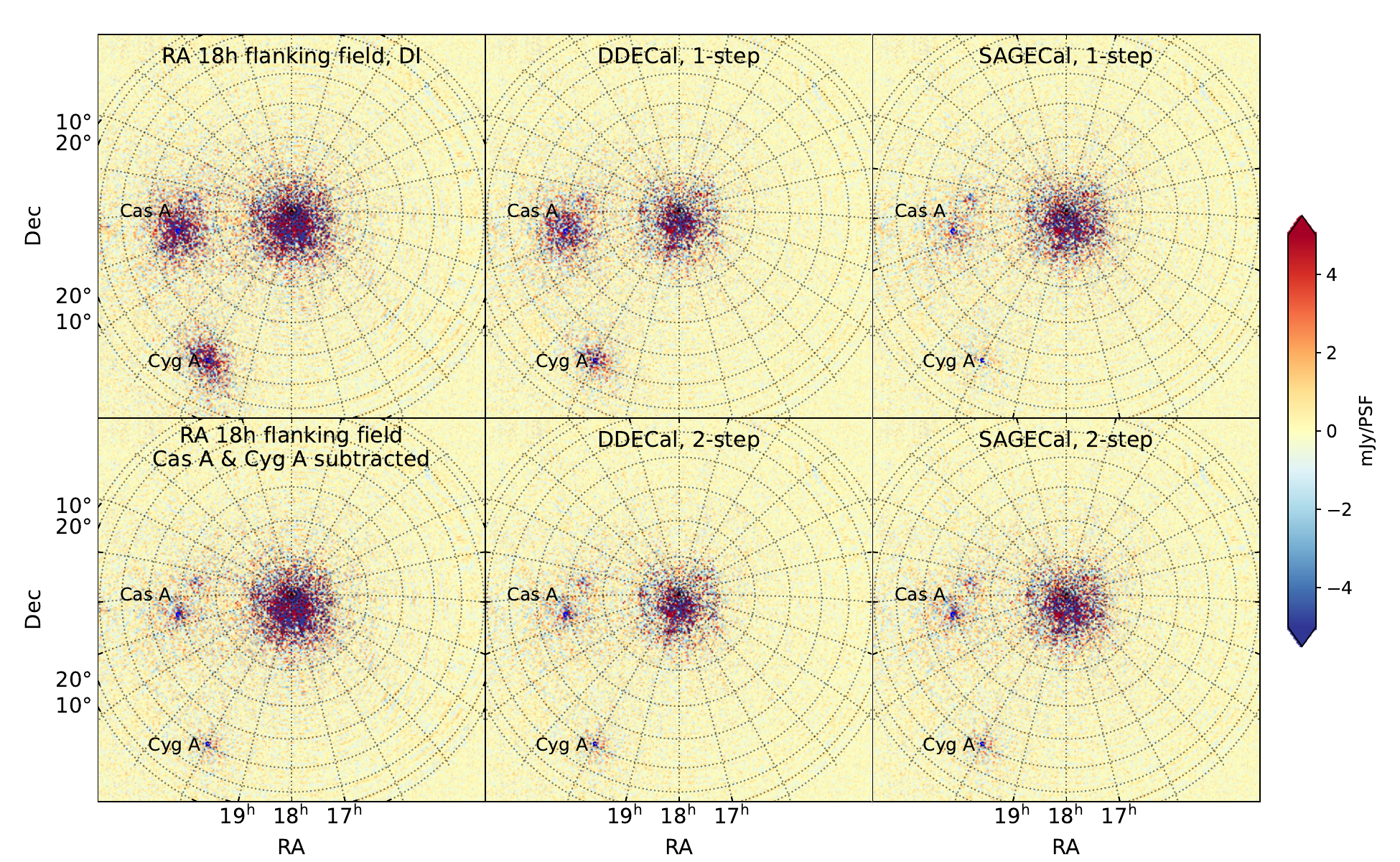}
    \caption{Full sky ($120^\circ\times120^\circ$) Stokes-I residual images created by using 69 sub-bands and $50-300\lambda$ baseline cut and integrating the full observation after DI-calibration, Cas~A and Cyg~A subtraction and DD-calibration with four different strategies with \textsc{ddecal} and \textsc{sagecal} on the RA 18h flanking field. The top left image shows the residuals after DI-calibration (top left). After DI-calibration, we subtract all sources including Cas~A and Cyg~A in one step (i.e. the 1-step method, middle and right on top), or we first subtract Cas~A and Cyg~A (i.e., the 2-step method, bottom left) and subtract sources in the centre (middle and right on the bottom). The flux of Cas~A and Cyg~A is largely reduced after Cas~A and Cyg~A subtraction (bottom left), while sources in the centre remain. The 1-step method with \textsc{sagecal} shows comparable performance in Cas~A and Cyg~A subtraction (top right), while \textsc{ddecal} still shows a high level of residuals (top middle). }
    \label{fig:very_wide_sap005_dd}
\end{figure*}

Fig.~\ref{fig:very_wide_sap005_dd} shows the residual Stokes-I images after DI-calibration (top left), Cas~A and Cyg~A subtraction with \textsc{ddecal} (bottom left) and DD-calibration with the four scenarios (second and third columns) on the RA 18h flanking field. By comparing the two images in the first column, before and after the subtraction of Cas~A and Cyg~A, we find that this extra step taken by \textsc{ddecal} significantly reduces the power from Cas~A and Cyg~A without changing the power around the phase centre. The subtraction of Cas~A and Cyg~A is not as efficient with the 1-step \textsc{ddecal} method (second panel on top) in Fig.~\ref{fig:very_wide_sap005_dd}), showing more residuals from Cas~A and Cyg~A after the DD-calibration compared to the other three scenarios.

This difference in the performance of the subtraction of Cas~A and Cyg~A between the 1-step and 2-step methods is very evident in \textsc{ddecal} (second column in Fig.~\ref{fig:very_wide_sap005_dd}), but not as much in \textsc{sagecal} (last column in Fig.~\ref{fig:very_wide_sap005_dd}). It is still unclear why the 1-step method performs better in subtracting sources in the primary beam than the 2-step method for \textsc{ddecal} in this specific case; and whether the existence of distant and bright sources (such as Cas~A and Cyg~A, in this case) in the sky model improves the prediction of nearby sources within the field of view.

However, this different performance of \textsc{ddecal} between the 1-step and 2-step methods shows the importance of optimising parameters during the DD-calibration. With the same calibration algorithm and sky model, the calibration performance can be different, depending on the exact parameters we use and the order in which the directions are solved for.

The 1-step \textsc{sagecal} method shows the best subtraction of Cyg~A compared to others, leaving the lowest power in the image (second panel on bottom in Fig.~\ref{fig:very_wide_sap005_dd}). One of the major differences between \textsc{ddecal} and \textsc{sagecal} is the application of time and frequency smearing correction and this correction is only applied for \textsc{sagecal} in this work. The better Cas~A and Cyg~A subtraction of \textsc{sagecal} could be due to this smearing correction.

In the far field in Fig.~\ref{fig:very_wide_sap005_dd}, Cas~A and Cyg~A are by far, the most dominant sources of residuals, even after DD-calibration. This is in line with the previous study on sources of excess variance in the LOFAR-EoR 21-cm power spectra~\citep{Gan_2022}. The dominant imprint of Cas~A and Cyg~A maybe are contributors of the excess power in the wedge.

\subsubsection{Gain smoothness difference in \textsc{ddecal} and \textsc{sagecal}}
\label{subsub:gain_smoothness}
\begin{figure*}
    \centering
    \includegraphics[width=\textwidth]{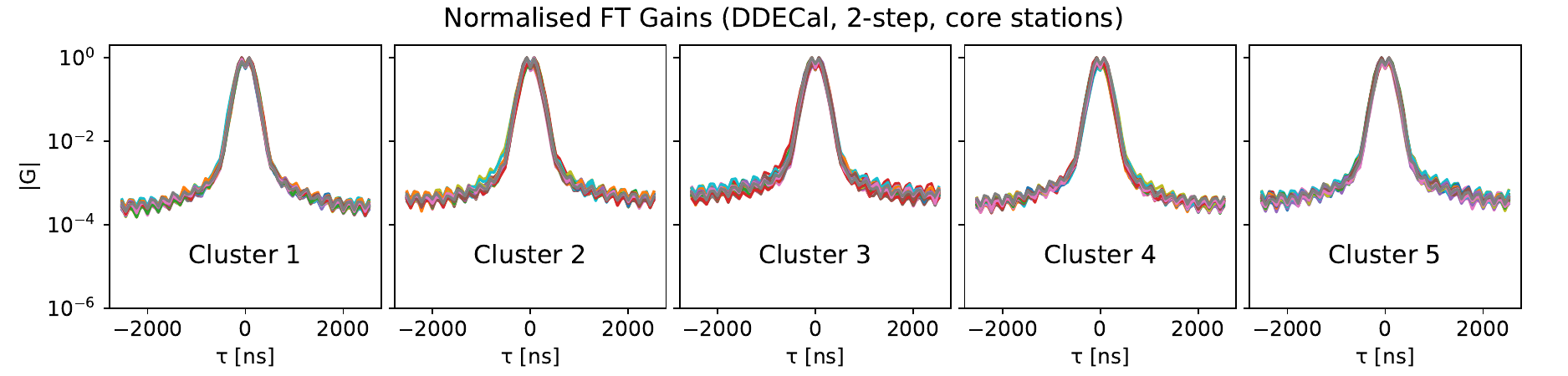}\\
    \includegraphics[width=\textwidth]{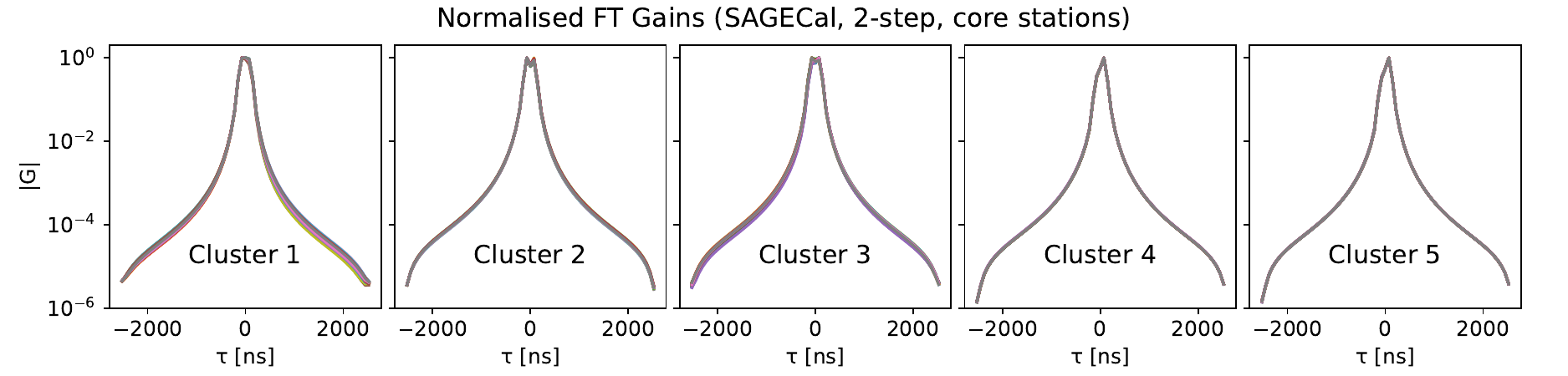}
    \caption{Peak-normalised gain solutions in delay space per station and cluster obtained by \textsc{ddecal} (on top) and \textsc{sagecal} (on bottom) with the 2-step method. Different colours denote solutions for different stations. Each polarisation component is added in quadrature. }
    \label{fig:ft_gains}
\end{figure*}
One of the main differences between the two DD-calibration algorithms is the implementation of frequency smoothness constraints. As discussed in section~\ref{sec:dd}, all sky signals, apart from the 21-cm signal are supposed to be smooth in frequency. By enforcing gains to be smooth in frequency, we can minimise signal suppression and avoid enhancing the noise variance introduced by calibration~\citep{Mevius_2022}. \textsc{ddecal} enforces this gain smoothness by convolving solutions with a Gaussian kernel of a given size for each iteration.

We test two different kernel sizes, 1 MHz and 4 MHz, of which the 4 MHz kernel turns out to be better for the analysis (i.e. better subtraction of the sky model). On the other hand, \textsc{sagecal} iteratively penalises solutions that deviate from a frequency smoothness prior by a quadratic term of a third-order Bernstein polynomial over the full bandwidth ($\sim13$ MHz in this case).

To understand the effects of different frequency constraints in \textsc{ddecal} and \textsc{sagecal}, we compare the delay $\tau$ transformed and peak-normalised (at $\tau=0$ ns) gains obtained by the two algorithms. 
Fig.~\ref{fig:ft_gains} shows the normalised gains obtained by \textsc{ddecal} (on top) and \textsc{sagecal} (on bottom) with the 2-step method for the first five clusters (from left to right) and core stations (in different colours) in delay space. For all clusters and stations, \textsc{sagecal} gain distributions show slightly narrower widths, compared to the ones from \textsc{ddecal}. A more noticeable difference is shown in the tails of gains at large delays. Gains from \textsc{ddecal} hit a noise floor ($|G|\sim10^{-4}$) at $|\tau| > 1000$ ns, while gains from \textsc{sagecal} continue to drop.
However, gains from \textsc{ddecal} and \textsc{sagecal} have similar distributions in delay space, despite the difference in the flux of the sky model and application of the beam model. We assume that the different frequency constraints used in the two algorithms have comparable effects in this analysis.

\subsection{Foreground removal: Gaussian process regression (GPR)}
\label{subsec:GPR}
In this subsection, we show the results of the GPR foreground removal after the four different DD-calibration scenarios, respectively.

The residual foregrounds after the DD-calibration can be further removed by GPR. GPR models each component in observation as a Gaussian Process (GP; \citealp{Rasmussen_2005}). The parametric GP model has five components: the foreground residuals that are composed of intrinsic sky emission and mode mixing contaminants related to the chromaticity of the instrument and calibration errors; the 21-cm signal; the spectrally uncorrelated noise; and the spectrally correlated excess noise.

\begin{table*}
      \caption[]{Summary of the GP model for each DD-calibration case. }
         \label{tab:gp_model}
    \centering
\begin{tabular}{l c c c c c}     
\hline
 Hyperparameter & Prior & \multicolumn{4}{c}{Estimate}  \\
   &  & \textsc{ddecal}, 1-step & \textsc{ddecal}, 2-step & \textsc{sagecal}, 1-step & \textsc{sagecal}, 2-step  \\
\hline
 $\eta_\text{sky}$ & $+\infty$ & - & - & - & -\\
 $\sigma^2_\text{sky}/\sigma^2_\text{n}$ & - & 355.7 & 341.5 & 553.7 & 502.9 \\
 $l_\text{sky}$ [MHz] &$\mathcal{U}(10.0,100.0)^*$ &  85.67 & 82.81 & 41.65 & 39.08 \\ \hline

 $\eta_\text{mix}$ & 3/2 & - & - & - & -\\
 $\sigma^2_\text{mix}/\sigma^2_\text{n}$ & - & 43.0 & 40.9 & 104.4 & 97.8  \\
 $l_\text{mix}$ [MHz] & $\mathcal{U}(0.5,20.0)$ &  3.342 & 3.183 & 3.697 & 3.686  \\ \hline

 $\eta_\text{ex}$ & 5/2 & - & - & - & -\\
 $\sigma^2_\text{ex}/\sigma^2_\text{n}$ & - & 6.5 & 5.5 & 7.0 & 5.9 \\
 $l_\text{ex}$ [MHz] & $\mathcal{U}(0.2,0.7)$ &  0.262 & 0.253 & 0.267 & 0.242  \\ \hline

 $\eta_\text{21}$ & 1/2 & - & - & - & -\\
 $\sigma^2_\text{21}/\sigma^2_\text{n}$ & - & 8.44E-04 & 1.46E-05 & 1.53E-06 & 3.21E-07 \\
 $l_\text{21}$ [MHz] & {$\mathcal{U}(0.1,1.2)$} &  0.806 &  0.808 & 0.808 & 0.810 \\
\hline
\end{tabular}
\tablefoot{$^\text{*} \mathcal{U}$ indicates a uniform prior.}
\end{table*}
The modelled GP components are summarised for the four different DD-calibration scenarios in Table.~\ref{tab:gp_model}. We refer readers to~\citet{Mertens_2020} for the detailed selection of the covariance model. GPR uses a Matern covariance function for modelling different components in the residuals. A Matern covariance function is defined with three hyperparameters, $\eta$, $\sigma$ and $l$. The parameter $\eta$ constrains the smoothness of the function, $\sigma^2$ is the variance and $l$ is the spectral coherence scale of each component. \citet{Mertens_2018} have found the most probable setting of GP and hyperparameter priors and the found values are used in this work.

We note that the coherence scales $l$ converge to similar values, especially, for the excess noise and 21-cm signal components, i.e., $l_\text{ex}$ and $l_\text{21}$, given the four different DD-calibration scenarios. This is expected because the four scenarios are based on the same observation and the DD-calibration step should not bias the excess variance and 21-cm signal.\footnote{The excess variance by definition is the extra noise that is above the thermal noise level and cannot be removed easily with DD-calibration or GPR~\citep{Mertens_2018,Gan_2022}.} On the other hand, the intrinsic sky and mix coherence scales, i.e., $l_\text{sky}$ and $l_\text{mix}$, depend on the calibration method in Table.~\ref{tab:gp_model}. For instant, \textsc{ddecal} shows longer intrinsic sky coherence scales (i.e., 82-85 MHz) compared to \textsc{sagecal} (i.e., 39-41 MHz).

This again can be explained by the difference in the residuals after the application of  \textsc{ddecal} and \textsc{sagecal}. As we discussed earlier, due to the application of the beam model, \textsc{ddecal} and \textsc{sagecal} end up with different residuals in the primary beam after DD-calibration. The sky and mix components are used to model these residuals. Hence it is reasonable to assume that the estimated parameters of the same calibration algorithm converge to similar values for the sky and mix components.

\subsection{Power spectra}
\label{subsec:PS}
\begin{figure*}
    \centering
    \includegraphics[width=\textwidth]{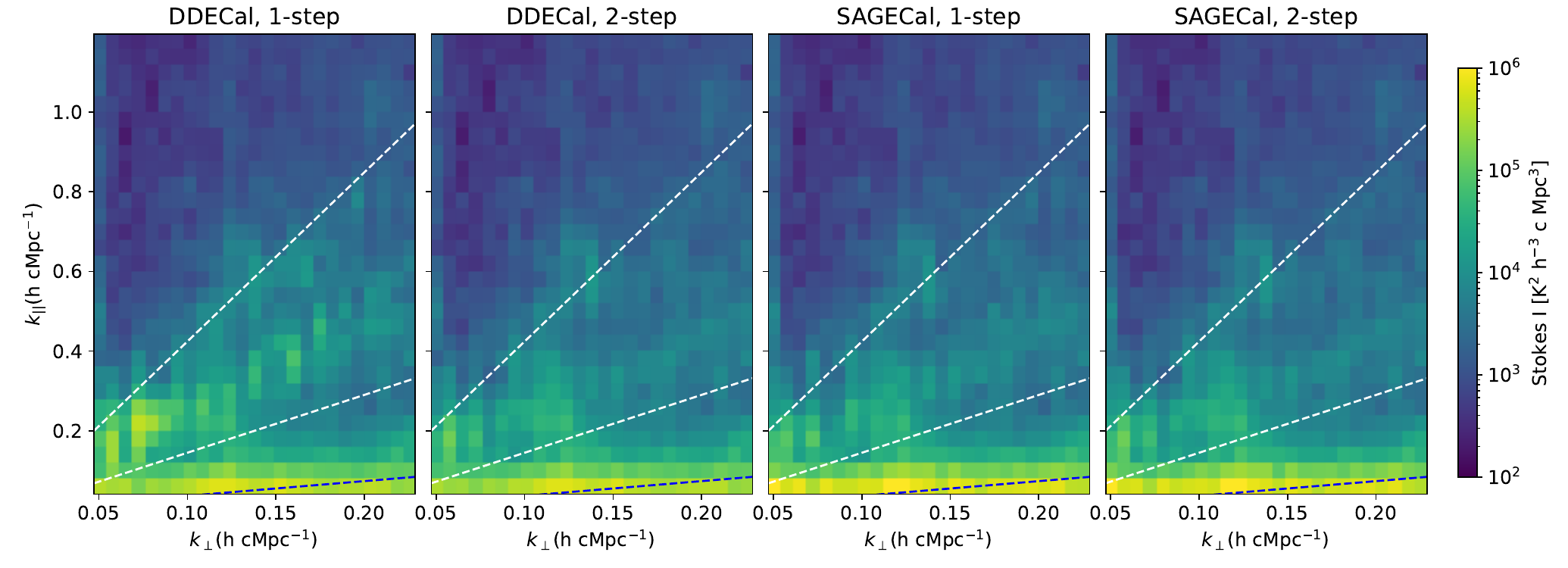}\\
    \includegraphics[width=\textwidth]{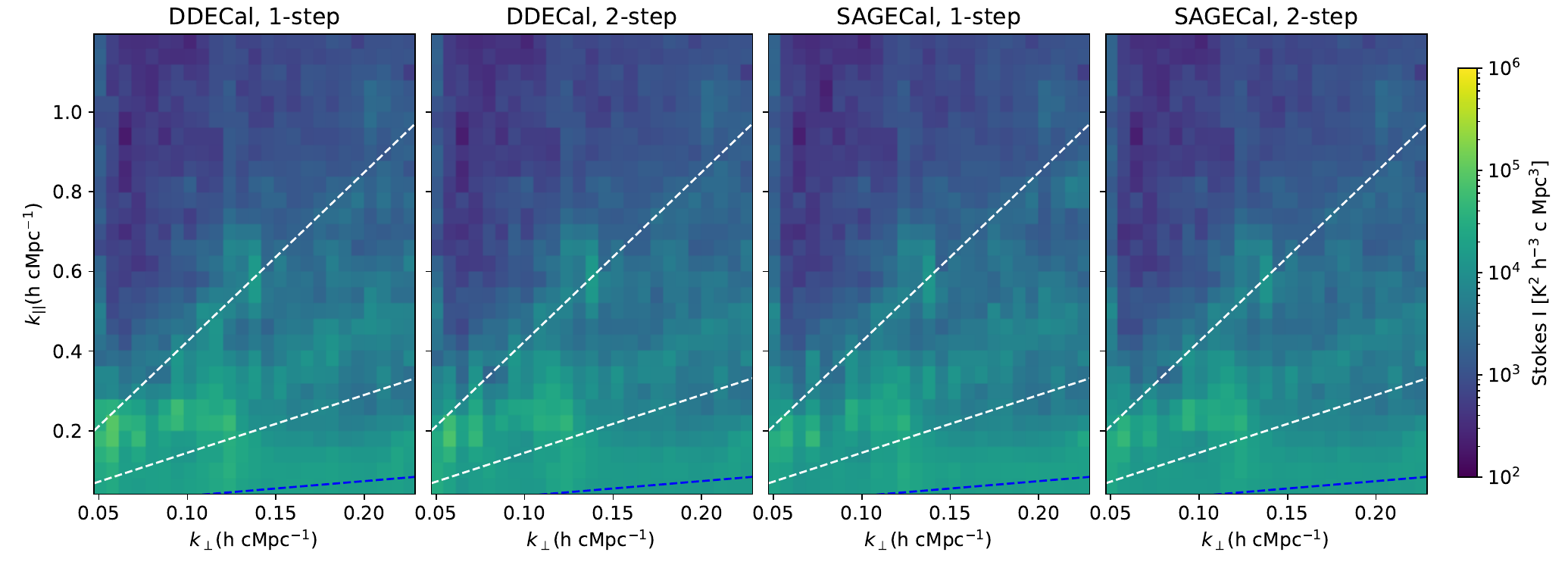}
    \caption{Cylindrical Stokes-I power spectra after DD-calibration (top) and GPR foreground removal (bottom) with four DD-calibration scenarios using \textsc{ddecal} and \textsc{sagecal}. The dashed lines, from bottom to top, correspond to the $5^\circ$ (the primary beam), $20^\circ$ and $90^\circ$ (instrumental horizon) delay lines from the phase centre. From the top row to the bottom row, the GPR foreground removal technique efficiently removes the residual power in the primary beam region. As we noted in the residual images in Fig.~\ref{fig:DD_images} (middle and bottom), \textsc{sagecal} (last two panels on top) leaves slightly higher power in the primary beam region than \textsc{ddecal} (first two panels on top) after DD-calibration. However, this difference in the primary beam disappears after the application of GPR (bottom). Also, the power spectrum of \textsc{ddecal} and 1-step method before GPR (first panel on top) has higher residual power between $20^\circ$ and $90^\circ$ delay lines compared to the rest, which indicates poor subtraction of Cas~A and Cyg~A. However, after GPR, the residual Stokes-I power spectra from the four different DD-calibration scenarios show very similar results.}
    \label{fig:2dps_DD_GPR}
\end{figure*}

\begin{figure}
    \centering
    \includegraphics[width=\columnwidth]{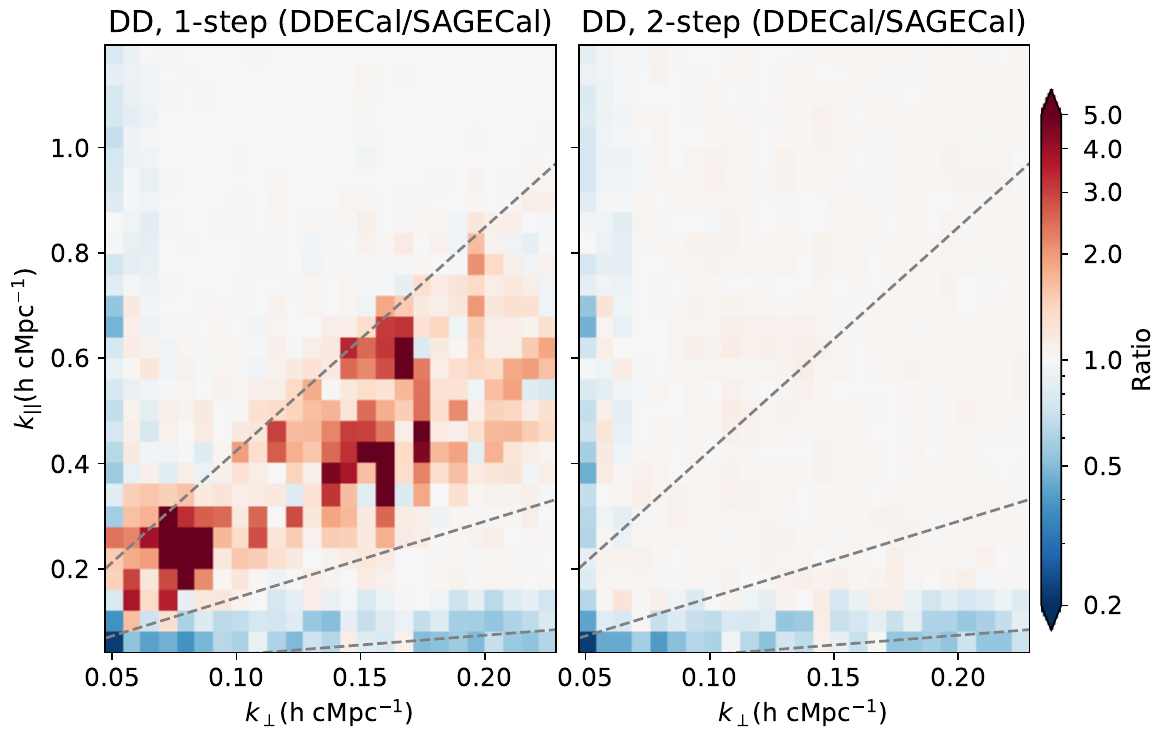}
    \includegraphics[width=\columnwidth]{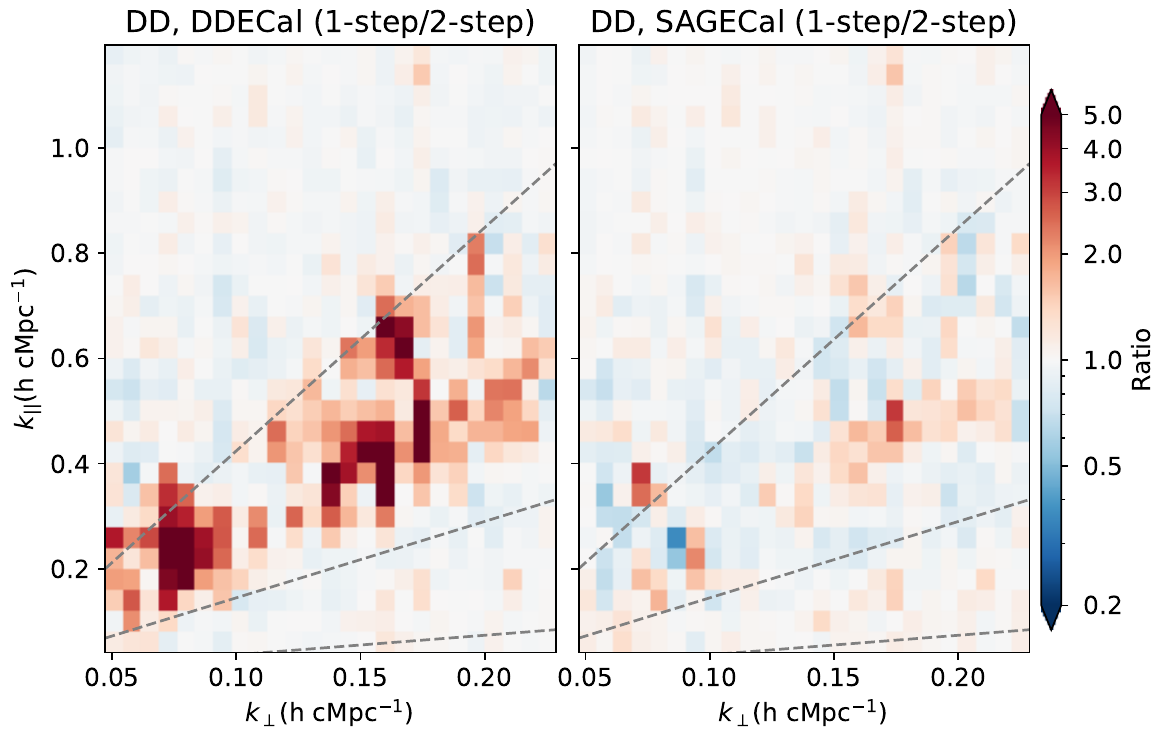}
    \caption{Cylindrical Stokes-I power spectra ratio after DD-calibration between \textsc{ddecal} and \textsc{sagecal} given a calibration strategy, the 1-step or 2-step method (on top), and between the 1-step and 2-step methods given a DD-calibration algorithm, \textsc{ddecal} or \textsc{sagecal} (on bottom). The dashed lines indicate the $5^\circ$ (the primary beam), $20^\circ$ and $90^\circ$ (instrumental horizon) delay lines from the phase centre from bottom to top. On top, red indicates excess power from \textsc{ddecal} and blue indicates excess power from \textsc{sagecal}. On bottom, red indicates excess power from the 1-step method and blue indicates excess power from the 2-step method.}
    \label{fig:2dps_dd_ratio}
\end{figure}

Fig.~\ref{fig:2dps_DD_GPR} shows the resulting cylindrical power spectra after the DD-calibration and foreground removal with four DD-calibration scenarios using \textsc{ddecal} and \textsc{sagecal}. Fig.~\ref{fig:2dps_dd_ratio} shows the cylindrical Stokes-I power-spectra ratio after DD-calibration between \textsc{ddecal} and \textsc{sagecal} for the 1-step or 2-step method (top), and the ratio between the 1-step and 2-step calibration method for the fixed DD-calibration algorithm (bottom). On top, red indicates excess power from \textsc{ddecal}, while blue indicates excess power from \textsc{sagecal}. On bottom, red indicates excess power from the 1-step method, and blue from the 2-step method. If the colour is close to white, it means that the ratio between the two methods is close to 1 and the difference is marginal. The dashed lines, from bottom to top, indicate the $5^\circ$ (the primary beam), $20^\circ$ and $90^\circ$ (instrumental horizon) delay lines from the phase centre.

The major difference between \textsc{ddecal} and \textsc{sagecal} is seen in the region of the primary beam (for both 1-step and 2-step methods in Fig.~\ref{fig:2dps_DD_GPR} on top and in Fig.~\ref{fig:2dps_dd_ratio} on top). We notice that the power in the primary beam region is lower when calibrated by \textsc{ddecal}. \textsc{sagecal} subtracts Cas~A and Cyg~A better than \textsc{ddecal} when the 1-step method is used (top left in Fig.~\ref{fig:2dps_dd_ratio}), however, the difference disappears when the 2-step method is used (top right in Fig.~\ref{fig:2dps_dd_ratio}). This different performance between the 1-step and 2-step methods with \textsc{ddecal} is also reflected on the bottom left in Fig.~\ref{fig:2dps_dd_ratio}. \textsc{ddecal} with the 1-step method shows significantly higher power between the $20^\circ$ and $90^\circ$ delay lines compared to the 2-step method. The difference between the 1-step and 2-step methods is, however, marginal for \textsc{sagecal} (bottom right in Fig.~\ref{fig:2dps_dd_ratio}).

After the GPR foreground removal, which is well suited to remove foreground in the primary beam region, the difference in the primary beam region between \textsc{ddecal} and \textsc{sagecal} is significantly reduced as shown in the bottom row of Fig.~\ref{fig:2dps_DD_GPR}.

\subsection{The North Celestial Pole results}
\label{subsec:NCP}
In this subsection, we present the DD-calibration results on the NCP processed by the standard LOFAR-EoR pipeline. The main differences between the DD-calibration examined in this work and the standard pipeline are summarised in Table.\ref{tab:dd_differences}. The NCP sky images after DI- and DD-calibration can be found in Appendix.\ref{app:diff_images}.

\begin{table*}[]
    \caption{Main differences between the DD-calibrations used in this work and in the standard LOFAR-EoR pipeline.}
        \centering
    \begin{tabular}{l|c|c}
    \hline
        Parameter & NCP & RA 18h flanking field\\\hline \hline
        Number of components & $\sim28000$ & 3389$^{*}$\\ \hline
        Number of clusters & 122 & 20* \\ \hline
        Baselines &  \multicolumn{2}{c}{$50-250\lambda$} \\ \hline
        Solution time interval & $2.5-20$ min$^{**}$ & 10 min\\ \hline
        Solution frequency interval & \multicolumn{2}{c}{per sub-band (195.3 kHz)}\\ \hline
    \end{tabular}
    \tablefoot{$^\text{*}$Without Cas~A and Cyg~A. $^\text{**}$Gain solution time interval varies, depending on the cluster in the NCP analysis.}
    \label{tab:dd_differences}
\end{table*}

\subsubsection{Images}

\begin{figure*}
    \centering
    \includegraphics[width=\textwidth]{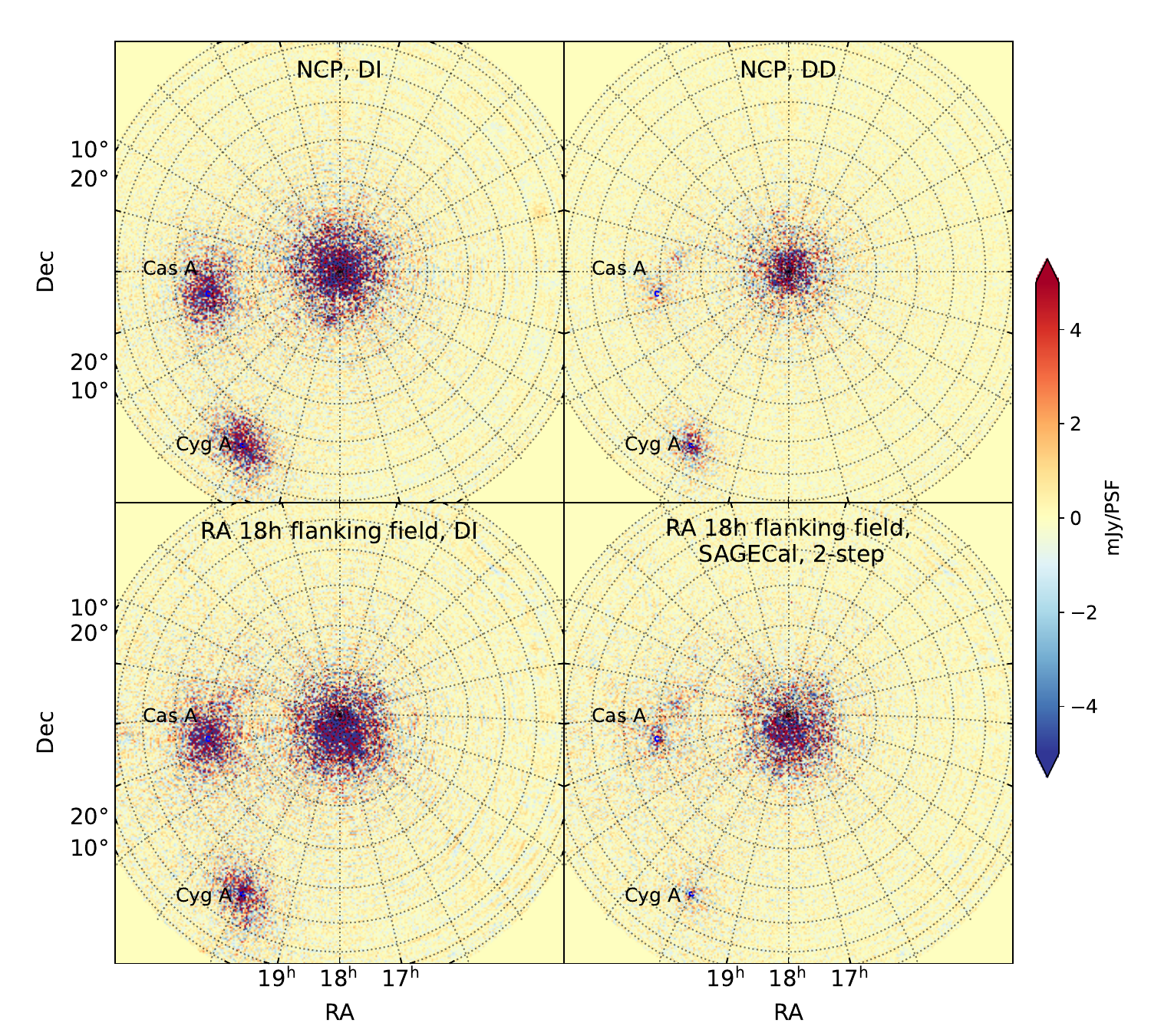}
    \caption{Comparison of full sky ($120^\circ\times120^\circ$) Stokes-I residual images created by using 69 sub-bands and $50-300\lambda$ baseline cut and integrating the full observation after DI (first column) and DD-calibration (second column) on the NCP (top) and on the RA 18h flanking field (bottom). The DD-calibration is performed by \textsc{sagecal} with an extensive sky model ($\sim28000$ sources) on the NCP and with a simpler sky model ($\sim3400$ sources) on the RA 18h flanking field. The residual power around the centre is substantially lower on the NCP (top right) than the flanking field (bottom right), due to using an extensive sky model with more directions during the DD-calibration. The power from Cas~A and Cyg~A is significantly reduced after DD-calibration. The residuals of Cas~A are stronger on the NCP (top right) than on the flanking field (bottom right). Unphysical sources below the horizon are masked. }
    \label{fig:very_wide_sap000_dd}
\end{figure*}

Fig.~\ref{fig:very_wide_sap000_dd} shows the full sky Stokes-I residual images on the NCP after the DI (top left) and DD-calibration (top right). The residual images after DI- and DD-calibration (with \textsc{sagecal} and 2-step method) are shown on bottom for comparison. Most sources show significantly reduced power after the DD-calibration on the NCP, however, we still see the imprint of Cas~A and Cyg~A in the residuals.

Compared to the residuals of the RA 18h flanking field in Fig.~\ref{fig:very_wide_sap005_dd} (bottom right), the residual power around the phase centre is significantly lower on the NCP, due to the application of the extensive sky model and more directions during the DD-calibration in the NCP processing.

However, the subtraction of Cas~A and Cyg~A does not show better performance on the NCP compared to the RA 18h flanking field. While the residuals of Cas~A are slightly more compact on the NCP than the RA 18h flanking field, the residuals of Cyg~A have significantly lower power on the RA 18h flanking field.

In Table.\ref{tab:ccsub_differences}, the calibration setup details for the Cas~A and Cyg~A subtraction on the NCP with the standard processing and flanking field with the best results using \textsc{sagecal} and the 1-step method are summarised. In both cases, the calibration is carried out by \textsc{sagecal} using the shapelet model. One main difference is the time interval of solutions in the two fields. The NCP uses a higher resolution, one solution per 2.5 min, compared to the 10-min interval of the RA 18h flanking field in this particular case. This also indicates that solving gains for a finer time interval does not always improve the calibration performance, because it is more likely to overfit the data and increase noise. Hence, finding an optimal calibration setup is crucial for the calibration performance, given a calibration algorithm and a sky model.

\begin{table}[]
    \caption{Comparison of the Cas~A and Cyg~A subtraction setup of the NCP in the standard processing with \textsc{sagecal} and the RA 18h flanking field with \textsc{sagecal} and the 1-step method.}
        \centering
    \begin{tabular}{l|c|c}
    \hline
        Parameter & NCP & RA 18h flanking field\\\hline \hline
        Calibration algorithm & \multicolumn{2}{c}{\textsc{sagecal}} \\ \hline
        Model & \multicolumn{2}{c}{Shapelet model} \\ \hline
        Number of clusters &  \multicolumn{2}{c}{2} \\ \hline
        Solution time interval & 2.5 min & 10 min\\ \hline
        Solution frequency interval & \multicolumn{2}{c}{per sub-band (195.3 kHz)}\\ \hline
    \end{tabular}
    \label{tab:ccsub_differences}
\end{table}

\subsubsection{Power spectra}

\begin{figure}
    \centering
    \includegraphics[width=\columnwidth]{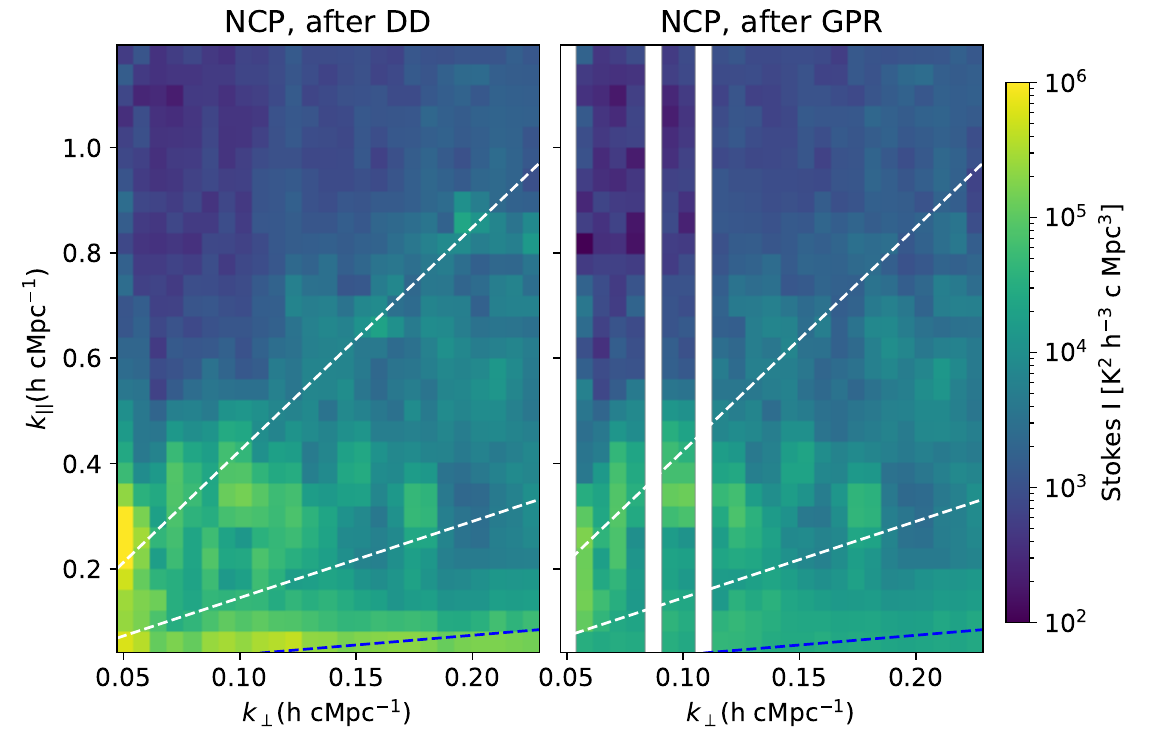}
    \caption{Cylindrical Stokes-I power spectra after DD-calibration (left) processed by the standard LOFAR-EoR pipeline with \textsc{sagecal} and GPR foreground removal (right) of a single observation night on the NCP. The dashed lines indicate the $5^\circ$ (the primary beam), $20^\circ$ and $90^\circ$ (instrumental horizon) delay lines from the phase centre from bottom to top, respectively. Some bad quality data are flagged on right. }
    \label{fig:2dps_ncp}
\end{figure}

Fig.~\ref{fig:2dps_ncp} shows the cylindrical Stokes-I power spectra after the DD-calibration and GPR foreground removal on the NCP. A few $k_\perp$ modes are flagged, due to bad data quality. Compared to the cylindrical Stokes-I residual power spectra of the RA 18h flanking field in Fig.~\ref{fig:2dps_DD_GPR} on bottom, the NCP has higher power on short baselines ($k_\perp\sim0.05-0.13$ h$\cdot$cMpc$^{-1}$), in particular, between the $20^\circ$ and $90^\circ$ delay lines and around the $90^\circ$ delay line.

\begin{figure*}
    \centering
    \includegraphics[trim={0 10 0 0},clip,width=\textwidth]{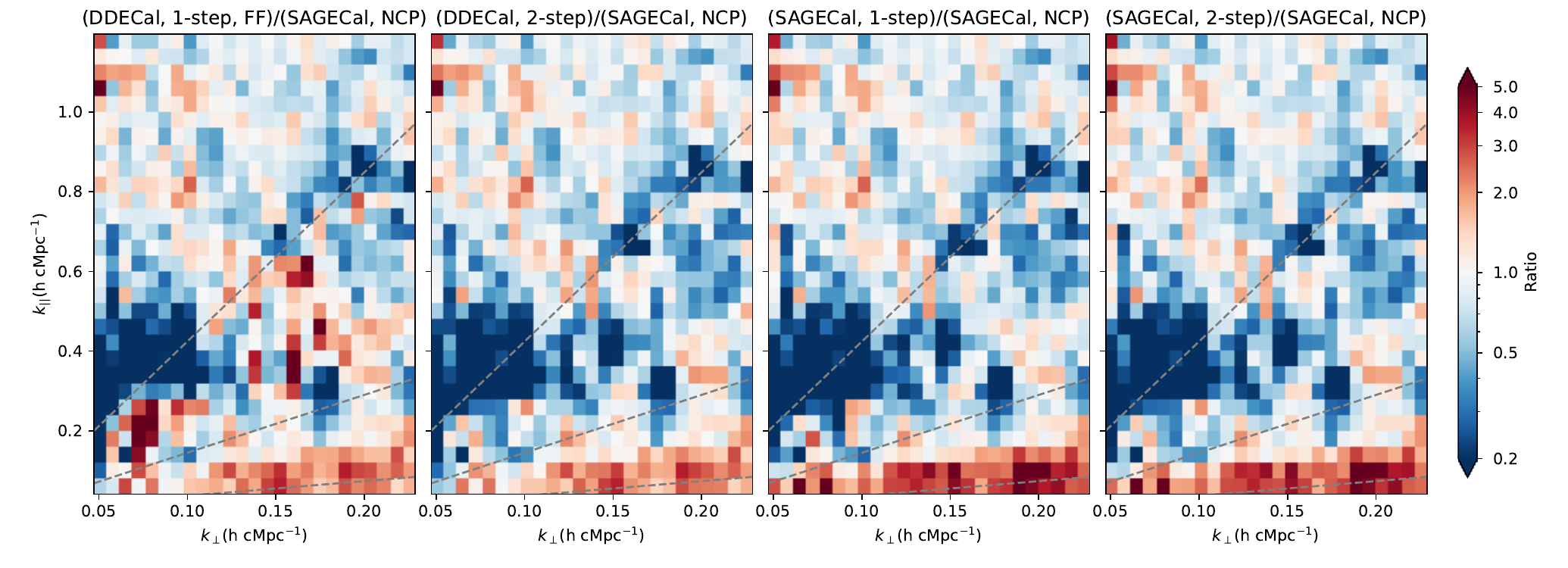}
    \includegraphics[trim={0 0 0 20},clip,width=\textwidth]{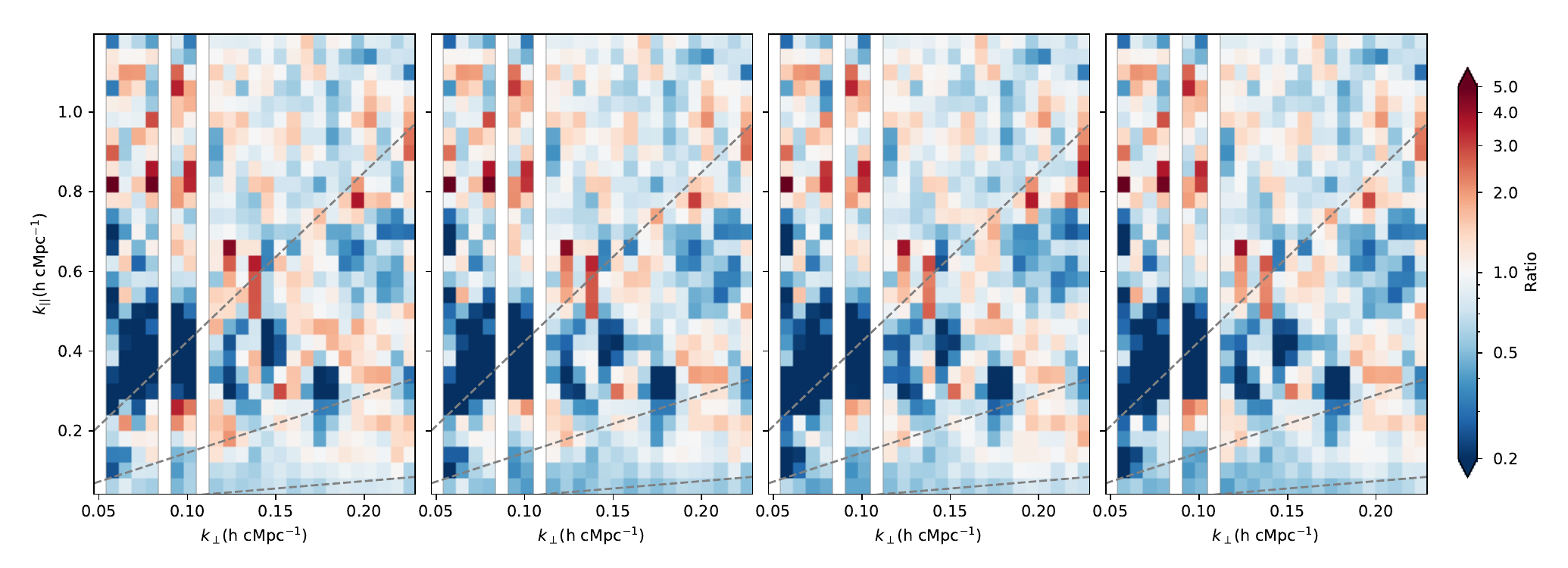}
    \caption{Cylindrical Stokes-I power spectra ratio after DD-calibration (top) and GPR foreground removal (bottom) between the flanking field and the NCP field. The dashed lines, from bottom to top, correspond to the $5^\circ$ (the primary beam), $20^\circ$ and $90^\circ$ (instrumental horizon) delay lines from the phase centre. Red indicates excess power from the flanking field and blue indicates excess power from the NCP field. Some bad data are flagged on bottom. From top to bottom, the residual foregrounds below the $20^\circ$ delay line (i.e. excess power from the flanking field) are well removed after GPR foreground removal.}
    \label{fig:2dps_ff_ncp}
\end{figure*}

Fig.~\ref{fig:2dps_ff_ncp} shows the ratio of cylindrical Stokes-I power spectra between the flanking field and NCP field after DD-calibration (top) and after GPR foreground removal (bottom). Red indicates excess power from the flanking field and blue indicates excess power from the NCP field. Again, a few $k_\perp$ modes are flagged, due to bad data quality after GPR foreground removal (bottom). As we have noticed before, the flanking field has higher residual power in the foreground region (below the $20^\circ$ delay line) after DD-calibration on top. This power is stronger with \textsc{sagecal} (last two panels on top) than with \textsc{ddecal} (first two panels on top). However, after GPR foreground removal, the excess power is well removed (bottom). Finally, the major difference comes from the excess power from the NCP on short baselines ($k_\perp\sim0.05-0.13$ h$\cdot$cMpc$^{-1}$) around the $90^\circ$ delay line as we have already seen in Fig.~\ref{fig:2dps_ncp}.

It is unclear where this extra power is from. Because the data on the two fields are from the same observing run only with the station beams and array phased up differently, they are supposed to have similar RFI and systematics. The NCP has a known disadvantage that stationary RFI sources can add coherently via its side-lobes. We, therefore, perform an extra RFI flagging step after DD-calibration to mitigate this effect on the NCP. However, this extra flagging step does not show a significant improvement.

\subsection{Spherically-averaged power spectra and upper limits}
\label{subsec:3dps}
\begin{figure*}
    \centering
    \includegraphics[width=0.7\textwidth]{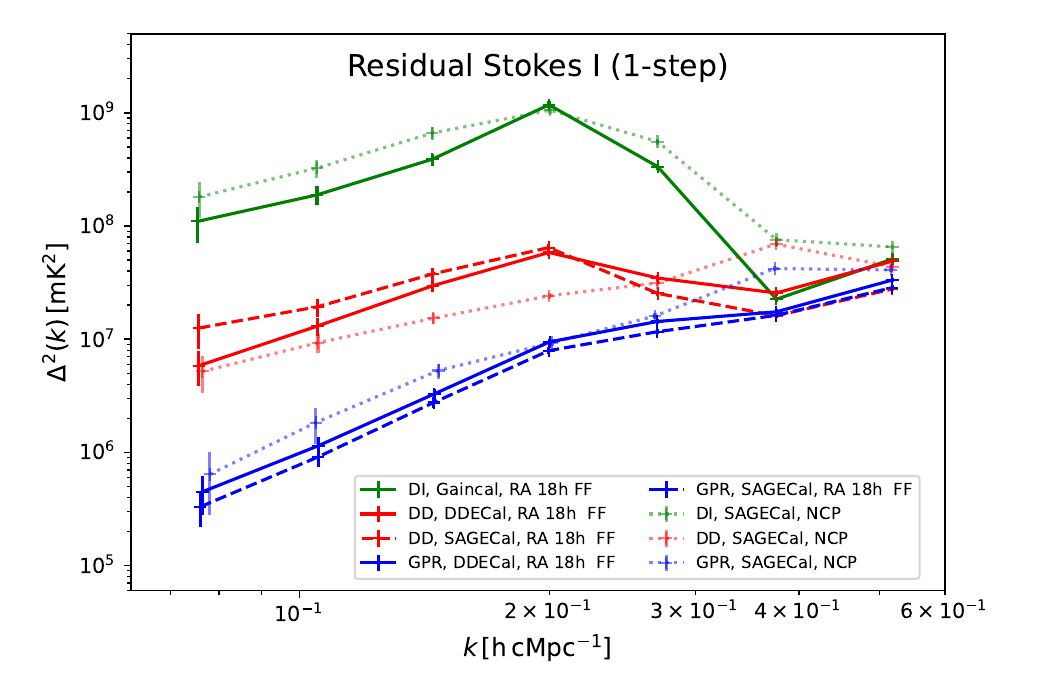}\\
    \includegraphics[width=0.7\textwidth]{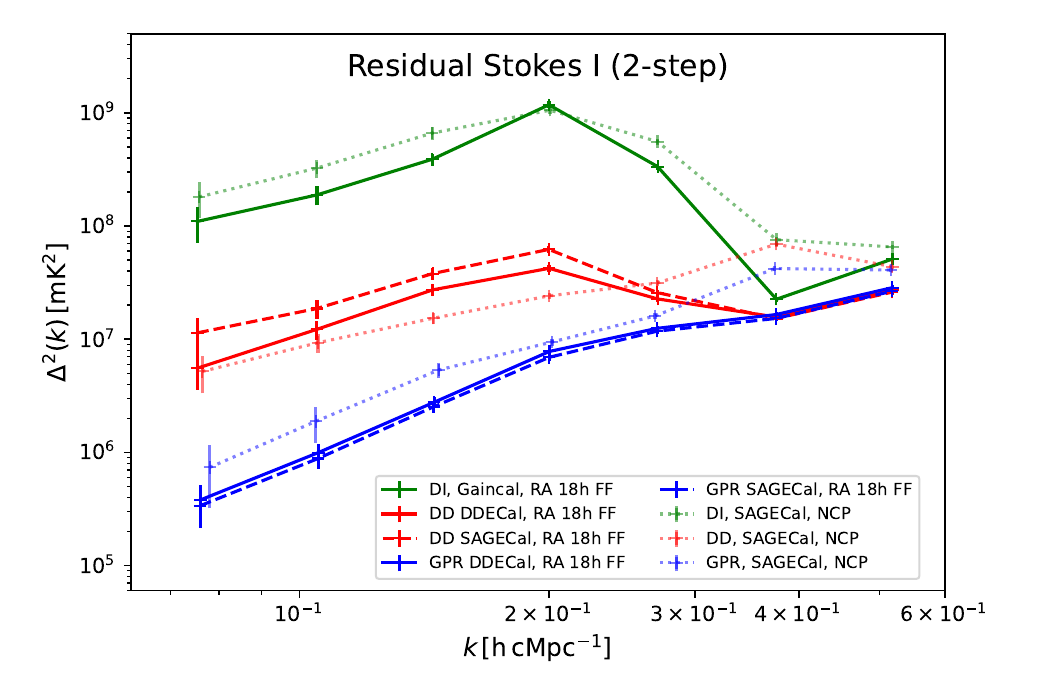}
    \caption{Spherically-averaged Stokes-I power spectra on the NCP calibrated by the standard LOFAR-EoR pipeline with \textsc{sagecal} (dotted lines) and the RA 18h flanking field calibrated by \textsc{ddecal} (solid lines) or \textsc{sagecal} (dashed lines) with the 1-step (on top) or 2-step (on bottom) method. Colours denote different processing stages. Green, red and blue denote the Stokes-I power spectra after DI-calibration, DD-calibration and GPR foreground removal, respectively. After each calibration stage, the Stokes-I power is reduced significantly. The NCP results show lower power after DD-calibration at low $k$ (< 0.3$\text{ h}\cdot\text{c}\text{Mpc}^{-1}$) compared to the RA 18h flanking field (in red), due to using an extensive sky model, however, after the GPR foreground removal, the RA 18h flanking field has lower power compared to the NCP (in blue). \textsc{ddecal} (solid lines) shows better subtraction of sources after DD-calibration (in red) compared to \textsc{sagecal} (dashed lines) in both 1-step and 2-step methods, due to the application of the beam model, however, this advantage disappears after GPR (in blue).}
    \label{fig:3dps_sap000_sap005}
\end{figure*}

Fig.~\ref{fig:3dps_sap000_sap005} shows the spherically-averaged Stokes-I power spectra of the RA 18h flanking field with four different DD-calibration scenarios and the NCP at different stages of the processing, the DI-calibration, DD-calibration and foreground removal (in green, red and blue). After the DI-calibration, the Stokes-I power is higher in the RA 18h flanking field than in the NCP, however, this tendency reverses after the DD-calibration (up to $k\sim0.7\text{ h}\cdot\text{c}\text{Mpc}^{-1}$), as we discussed earlier, likely due to the extensive sky model and clustering used for the DD-calibration on the NCP. However, this advantage disappears once we apply the GPR foreground removal technique. The Stokes-I power of the RA 18h flanking field is lower than that of the NCP by a factor of $1.2-2$ over $k=0.075-0.6$ h$\cdot$cMpc$^{-1}$.

We also compare the 2-$\sigma$ upper limits of the 21-cm signal on the RA 18h flanking field and NCP (in Table.\ref{tab:upperlimits}). Compared to the NCP results, the RA 18h flanking field shows around $10-30\%$ improved upper limits over all $k$ values considered for the four different calibration scenarios.

The 2-$\sigma$ upper limits on the 21-cm signal after the calibration and foreground removal by the four different DD-calibration scenarios are summarised in Table.\ref{tab:upperlimits}. The 2-step \textsc{sagecal} method presents the best upper limit results compared to others. Within the four cases studied, \textsc{sagecal} provides $5-10\%$ better upper limits compared to \textsc{ddecal} given a calibration strategy, 1-step or 2-step. The 2-step method produces $3-8\%$ better upper limits given a calibration algorithm, \textsc{ddecal} or \textsc{sagecal}. This difference is minor and could also come from the different model assumptions (e.g. between apparent and intrinsic flux models, and/or between different Cas~A and Cyg~A models), apart from the difference in the model clustering and the application of the primary beam model.

\begin{table}
  \caption[]{2-$\sigma$ upper limit of the 21-cm signal $\Delta^{2}_\text{21}$ from a single observation night on the RA 18h flanking field with different DD-calibration scenarios and the NCP calibrated by the standard pipeline with \textsc{sagecal}. }
  \label{tab:upperlimits}
 \centering
\begin{minipage}{.44\linewidth}
$$
     \begin{array}{l r }
        \textsc{ddecal}\text{, 1-step} & \\
        \hline
        \noalign{\smallskip}
        k & \Delta^{2}_\text{21}\\
        $[h$\cdot$c$\text{Mpc}^{-1}$]$ & $[\text{mK}$^2$]$\\
        \noalign{\smallskip}
        \hline
        \noalign{\smallskip}
        0.0764 & (766)^2 \\
        0.1055 & (1154)^2 \\
        0.1454 & (1874)^2 \\
        0.2002 & (3102)^2 \\
        0.2699 & (3755)^2 \\
        0.3756 & (3968)^2 \\
        0.5184 & (5299)^2 \\
        \noalign{\smallskip}
        \hline
        \noalign{\smallskip}
        \noalign{\smallskip}
     \end{array}
$$
\end{minipage}
\begin{minipage}{.44\linewidth}
 $$
     \begin{array}{l r }
        \textsc{ddecal}\text{, 2-step} & \\
        \hline
        \noalign{\smallskip}
        k & \Delta^{2}_\text{21}\\
        $[h$\cdot$c$\text{Mpc}^{-1}$]$ & $[\text{mK}$^2$]$\\
        \noalign{\smallskip}
        \hline
        \noalign{\smallskip}
        0.0760 & (736)^2 \\
        0.1054 & (1067)^2 \\
        0.1454 & (1721)^2 \\
        0.2002 & (2817)^2 \\
        0.2699 & (3487)^2 \\
        0.3755 & (3843)^2 \\
        0.5183 & (4771)^2 \\
        \noalign{\smallskip}
        \hline
        \noalign{\smallskip}
        \noalign{\smallskip}
     \end{array}
 $$
\end{minipage}
\begin{minipage}{.44\linewidth}
 $$
     \begin{array}{l r }
        \textsc{sagecal}\text{, 1-step} & \\
        \hline
        \noalign{\smallskip}
        k & \Delta^{2}_\text{21}\\
        $[h$\cdot$c$\text{Mpc}^{-1}$]$ & $[\text{mK}$^2$]$\\
        \noalign{\smallskip}
        \hline
        \noalign{\smallskip}
        0.0759 & (678)^2 \\
        0.1054 & (1021)^2 \\
        0.1452 & (1725)^2 \\
        0.1999 & (2839)^2 \\
        0.2700 & (3380)^2 \\
        0.3755 & (3854)^2 \\
        0.5182 & (4892)^2 \\
        \noalign{\smallskip}
        \hline
        \noalign{\smallskip}
        \noalign{\smallskip}
     \end{array}
 $$
\end{minipage}
\begin{minipage}{.44\linewidth}
 $$
     \begin{array}{l r }
        \textsc{sagecal}\text{, 2-step} & \\
        \hline
        \noalign{\smallskip}
        k & \Delta^{2}_\text{21}\\
        $[h$\cdot$c$\text{Mpc}^{-1}$]$ & $[\text{mK}$^2$]$\\
        \noalign{\smallskip}
        \hline
        \noalign{\smallskip}
        0.0759 & (666)^2 \\
        0.1053 & (1007)^2 \\
        0.1453 & (1636)^2 \\
        0.2000 & (2660)^2 \\
        0.2699 & (3395)^2 \\
        0.3755 & (3688)^2 \\
        0.5181 & (4605)^2 \\
        \noalign{\smallskip}
        \hline
        \noalign{\smallskip}
        \noalign{\smallskip}
     \end{array}
 $$
\end{minipage}

 $$
     \begin{array}{l r }
        \text{NCP, standard} & \\
        \hline
        \noalign{\smallskip}
        k & \Delta^{2}_\text{21}\\
        $[h$\cdot$c$\text{Mpc}^{-1}$]$ & $[\text{mK}$^2$]$\\
        \noalign{\smallskip}
        \hline
        \noalign{\smallskip}
        0.0781 & (1041)^2 \\
        0.1047 & (1608)^2 \\
        0.1471 & (2457)^2 \\
        0.2016 & (3167)^2 \\
        0.2685 & (4041)^2 \\
        0.3740 & (6508)^2 \\
        0.5168 & (5896)^2 \\
        \noalign{\smallskip}
        \hline
        \noalign{\smallskip}
        \noalign{\smallskip}
     \end{array}
 $$
\end{table}

\section{Conclusions}
\label{sec:conclusions}
In this work, we have compared the performance of two DD-calibration algorithms, \textsc{ddecal} and \textsc{sagecal}, in the context of LOFAR-EoR 21-cm power spectra by processing a single observation night on a flanking field of the North Celestial Pole (NCP) obtained by Low Frequency Array (LOFAR). We applied two different strategies for subtracting the very bright sources Cas~A and Cyg~A, predicting and subtracting the two sources simultaneously with the sky model, namely, the 1-step method, or in a separate step before predicting and subtracting the sky model, the 2-step method. We conclude the following:

(1) We do notice that there are differences between the two DD-calibration algorithms. \textsc{ddecal} shows a better performance in subtracting sources in the primary beam region, probably due to the application of the beam model during the DD-calibration. This suggests that having a beam model during the DD-calibration significantly improves the calibration performance, especially, in the primary beam region.

(2) \textsc{sagecal}, on the other hand, shows a better performance in subtracting Cas~A and Cyg~A. While predicting and subtracting Cas~A and Cyg~A in a separate step does not change the DD-calibration results significantly for \textsc{sagecal}, it does make a significant difference for \textsc{ddecal}. The time and frequency smearing correction is applied for \textsc{sagecal} but not for \textsc{ddecal} in this work. The difference in subtracting Cas~A and Cyg~A could be due to the application of this smearing correction.

(3) The difference of the residual power in the primary beam region between \textsc{ddecal} and \textsc{sagecal} becomes marginal when the GPR foreground removal is applied after DD-calibration.

(4) We also compare the results on the RA 18h flanking field with the NCP results processed by the standard LOFAR-EoR pipeline. The standard processing pipeline uses a very extended sky model (with $\sim28000$ sources) and 122 directions for the DD-calibration which makes the processing computationally expensive.

(5) For the four different DD-calibration scenarios studied, comparable upper limits on the 21-cm power spectra on the NCP flanking field are achieved, using a simpler sky model (with $\sim3500$ sources including Cas~A and Cyg~A) and fewer directions (20 directions), when the foreground removal technique, Gaussian Process Regression (GPR), is used after DD-calibration.

(6) In both NCP and RA 18h flanking field results, even after DD-calibration, Cas~A and Cyg~A are the most dominant sources of residuals in the far field in full sky images in Fig.~\ref{fig:very_wide_sap000_dd} and~\ref{fig:very_wide_sap005_dd}, which agrees with the previous study on sources of excess variance in the LOFAR-EoR 21-cm power spectra~\citep{Gan_2022}. They may be contributors to the excess noise in the wedge.

Based on our analysis, we suggest the following strategies for future improvements:

\textbf{Apply time and frequency smearing corrections for \textsc{ddecal}}: the latest version of \textsc{ddecal} corrects for the time and frequency smearing. This correction is not applied during the DD-calibration process in this work. In future, we would like to include the smearing correction during the DD-calibration and investigate whether it further improves the calibration performance, especially the subtraction of Cas~A and Cyg~A.

\textbf{Apply a beam model for \textsc{sagecal}}: the future versions of \textsc{sagecal} will support the LOFAR beam model. Our results with \textsc{ddecal} show that applying a beam model is likely to improve the source subtraction around the phase centre substantially. We expect to achieve similar source subtraction performance with \textsc{sagecal}, once we apply a beam model for \textsc{sagecal}.

\textbf{Process flanking fields for cross-checks}: in this work, we have processed a single observation night on two different fields, the NCP and one of its flanking fields, using different calibration setups and compared their results. While the data sets from the same observation share the same or very similar RFI, ionosphere and systematics, we note that the residuals in the two fields look rather different than expected. In particular, the NCP shows extra power above the wedge on short baselines that is not present in the RA 18h flanking field. We suspect that this power partially comes from the residuals of Cas~A and Cyg~A, however, more investigations are needed to clarify the source(s) of the extra power. By processing other flanking fields of the same observation and imaging ground planes, we will be able to identify whether this is a particular problem of the NCP field that is related to the beam or a calibration issue.

\textbf{Optimise DD-calibration parameters}: by comparing the performance of the four different DD-calibration scenarios, even with the same sky model, depending on the calibration parameters and strategies we use, such as frequency smoothing constraints or the number of clusters for particularly bright sources, the calibration outcome can be significantly different. While there are some studies on the regularisation of frequency in the DD-calibration~\citep{Yatawatta_2015,Yatawatta_2016fine,Mevius_2022}, more studies are needed, in particular, for the selection of the number of clusters and solution intervals. Different calibration parameters can be tested relatively straightforward because it does not require modifying the existing sky model or pipeline.

\begin{acknowledgements}
      HG and LVEK would like to acknowledge support from the Centre for Data Science and Systems Complexity (DSSC) at the University of Groningen and a \emph{Marie Sk\l{}odowska-Curie COFUND} grant, no.754315. BKG and LVEK acknowledge the financial support from the European Research Council (ERC) under the European Union’s Horizon 2020 research and innovation programme (Grant agreement No. 884760, "CoDEX").
\end{acknowledgements}

\section*{Data Availability}

The data underlying this article will be shared upon request to the corresponding author.

\bibliographystyle{aa} 
\bibliography{bib} 

%

\begin{appendix} 

\section{NCP flanking field configuration}
In Table~\ref{tab:ff}, the pointing and beam number of the NCP and six flanking fields are summarised. The phase centres of the six flanking fields are located $4^\circ$ from the NCP field. In this work, we focus on the RA 18h field.

\begin{table*}
      \caption[]{Summary of the NCP flanking field positions. SAP stands for Sub-Array Pointing. }
         \label{tab:ff}
    \centering
\begin{tabular}{l c l}     
\hline
\noalign{\smallskip}
 Field & Beam number & Pointing (\textit{J2000.0}) \\
\noalign{\smallskip}
\hline
\noalign{\smallskip}
NCP & SAP000 & $00^\text{h}00^\text{m}00^\text{s  } $+90$^{\circ}00^{\prime}00^{\prime\prime}$\\
3C61.1 & SAP001 & $02^\text{h}00^\text{m}00^\text{s  }
$+86$^{\circ}00^{\prime}00^{\prime\prime}$\\
RA 6h flanking field & SAP002 & $06^\text{h}00^\text{m}00^\text{s  } $+86$^{\circ}00^{\prime}00^{\prime\prime}$\\
RA 10h flanking field & SAP003 & $10^\text{h}00^\text{m}00^\text{s  } $+86$^{\circ}00^{\prime}00^{\prime\prime}$\\
RA 14h flanking field &  SAP004 & $14^\text{h}00^\text{m}00^\text{s  } $+86$^{\circ}00^{\prime}00^{\prime\prime}$\\
RA 18h flanking field &  SAP005 & $18^\text{h}00^\text{m}00^\text{s  } $+86$^{\circ}00^{\prime}00^{\prime\prime}$\\
RA 22h flanking field &  SAP006 & $22^\text{h}00^\text{m}00^\text{s  } $+86$^{\circ}00^{\prime}00^{\prime\prime}$\\
\noalign{\smallskip}
\hline
\end{tabular}
\end{table*}

\section{Sky model details}
Here we provide detailed information about the sky model on the RA 18h flanking field. The flux of the model is scaled using four calibrators summarised in Table.~\ref{tab:flux_scaling}. The details of sky model clustering are summarised in Table.~\ref{tab:model}.

\begin{table*}
      \caption[]{Four bright radio sources around the position of RA 18h flanking field selected for setting the absolute flux of the sky model.}
         \label{tab:flux_scaling}
    \centering
\begin{tabular}{l l l l l}     
\hline
\noalign{\smallskip}
 Source & Position (\textit{J2000.0}) & Frequency & Peak flux & Reference\\
  & Ra, Dec & [MHz] & [Jy] & \\
\noalign{\smallskip}
\hline
\noalign{\smallskip}
 J190401.7+8536 & $19^\text{h}04^\text{m}03^\text{s  } $+85$^{\circ}36^{\prime}$ & 118.75 & $5.069\pm0.549$ & \cite{Zheng_2016ApJ...832..190Z}\\
 6C B184741+851139 & $18^\text{h}37^\text{m}12.220^\text{s  } $+85$^{\circ}14^{\prime}49.40^{\prime\prime} $ & 151.5 & $4.09\pm0.035$ & \cite{Baldwin_1985MNRAS.217..717B}\\
 6C B174711+844656 & $17^\text{h}37^\text{m}40.83^\text{s  } $+84$^{\circ}45^{\prime}43.9^{\prime\prime} $ & 151.5 & $4.56\pm0.035$ & \cite{Baldwin_1985MNRAS.217..717B}\\
 6C B163113+855559 & $16^\text{h}19^\text{m}40.62^\text{s  } $+85$^{\circ}49^{\prime}21.2^{\prime\prime} $ & 151.5 & $6.2\pm0.035$ & \cite{Baldwin_1985MNRAS.217..717B}\\
\noalign{\smallskip}
\hline
\end{tabular}
\end{table*}

\begin{table*}
      \caption[]{Summary of the \texttt{CLEAN} component model and its clustering.}
         \label{tab:model}
    \centering
\begin{tabular}{c l r r r r }     
\hline
 Cluster & Position (\textit{J2000.0}) & Number of & Maximum & Total flux& Maximum  \\
   & RA [hour], Dec [deg] & sources &  flux [Jy] & density [Jy] & separation [deg] \\
\hline
 1 & $13^\text{h}46^\text{m}06.446^\text{s  } $+85$^{\circ}30^{\prime}02.564^{\prime\prime}$ & 212 & 5.185 & 71.87 & 2.134 \\
 2 & $15^\text{h}38^\text{m}57.775^\text{s  } $+86$^{\circ}46^{\prime}08.711^{\prime\prime}$ & 205 & 6.400 & 61.51 & 1.343  \\
 3 & $16^\text{h}44^\text{m}56.439^\text{s  } $+82$^{\circ}00^{\prime}46.566^{\prime\prime}$ & 157 & 4.060 & 52.54 & 1.394  \\
 4 & $17^\text{h}46^\text{m}29.922^\text{s  } $+85$^{\circ}25^{\prime}39.948^{\prime\prime}$ & 218 & 5.876 & 49.19 & 1.268  \\
 5 & $19^\text{h}15^\text{m}00.497^\text{s  } $+84$^{\circ}26^{\prime}46.767^{\prime\prime}$ & 153 & 4.471 & 48.08 & 1.267  \\
 6 & $21^\text{h}04^\text{m}07.835^\text{s  } $+84$^{\circ}12^{\prime}30.181^{\prime\prime}$ & 152 & 8.257 & 47.69 & 1.570  \\
 7 & $22^\text{h}54^\text{m}00.975^\text{s  } $+88$^{\circ}23^{\prime}08.218^{\prime\prime}$ & 238 & 5.684 & 45.48 & 2.149  \\
 8 & $20^\text{h}01^\text{m}40.512^\text{s  } $+85$^{\circ}59^{\prime}54.030^{\prime\prime}$ & 184 & 6.044 & 43.34 & 1.397  \\
 9 & $15^\text{h}58^\text{m}32.933^\text{s  } $+84$^{\circ}33^{\prime}04.712^{\prime\prime}$ & 196 & 22.737 & 41.71 & 1.364  \\
 10 & $18^\text{h}26^\text{m}15.815^\text{s  } $+87$^{\circ}37^{\prime}23.340^{\prime\prime}$ & 220 & 3.209 & 38.07 & 1.502  \\
 11 & $12^\text{h}22^\text{m}01.146^\text{s  } $+88$^{\circ}28^{\prime}52.985^{\prime\prime}$ & 246 & 2.873 & 37.40 & 1.850  \\
 12 & $15^\text{h}55^\text{m}46.173^\text{s  } $+81$^{\circ}59^{\prime}38.825^{\prime\prime}$ & 147 &  5.261 & 35.62 & 1.776  \\
 13 & $20^\text{h}05^\text{m}39.918^\text{s  } $+83$^{\circ}06^{\prime}40.170^{\prime\prime}$ & 119 & 2.061 & 33.69 & 1.286  \\
 14 & $22^\text{h}23^\text{m}24.174^\text{s  } $+86$^{\circ}16^{\prime}28.896^{\prime\prime}$ & 186 & 5.828 & 29.83 & 2.521  \\
 15 & $19^\text{h}21^\text{m}26.177^\text{s  } $+82$^{\circ}13^{\prime}33.547^{\prime\prime}$ & 95 & 2.762 & 25.27 & 1.489  \\
 16 & $17^\text{h}39^\text{m}37.842^\text{s  } $+82$^{\circ}09^{\prime}27.905^{\prime\prime}$ & 106 & 2.050 & 22.13 & 1.053  \\
 17 & $18^\text{h}20^\text{m}56.808^\text{s  } $+83$^{\circ}20^{\prime}57.184^{\prime\prime}$ & 150 & 1.014 & 21.49 & 1.097  \\
 18 & $17^\text{h}04^\text{m}01.348^\text{s  } $+83$^{\circ}46^{\prime}54.388^{\prime\prime}$ & 155 & 2.962 & 21.04 & 1.173  \\
 19 & $15^\text{h}08^\text{m}22.540^\text{s  } $+83$^{\circ}36^{\prime}40.085^{\prime\prime}$ & 166 & 2.361 & 20.83 & 1.440  \\
 20 & $18^\text{h}31^\text{m}11.244^\text{s  } $+81$^{\circ}33^{\prime}30.857^{\prime\prime}$ & 84 & 2.125 & 19.51 & 1.250  \\
\hline
\end{tabular}
\end{table*}

\section{Difference images of DD-calibration residuals}
\begin{figure*}
    \centering
    \includegraphics[width=0.9\textwidth]{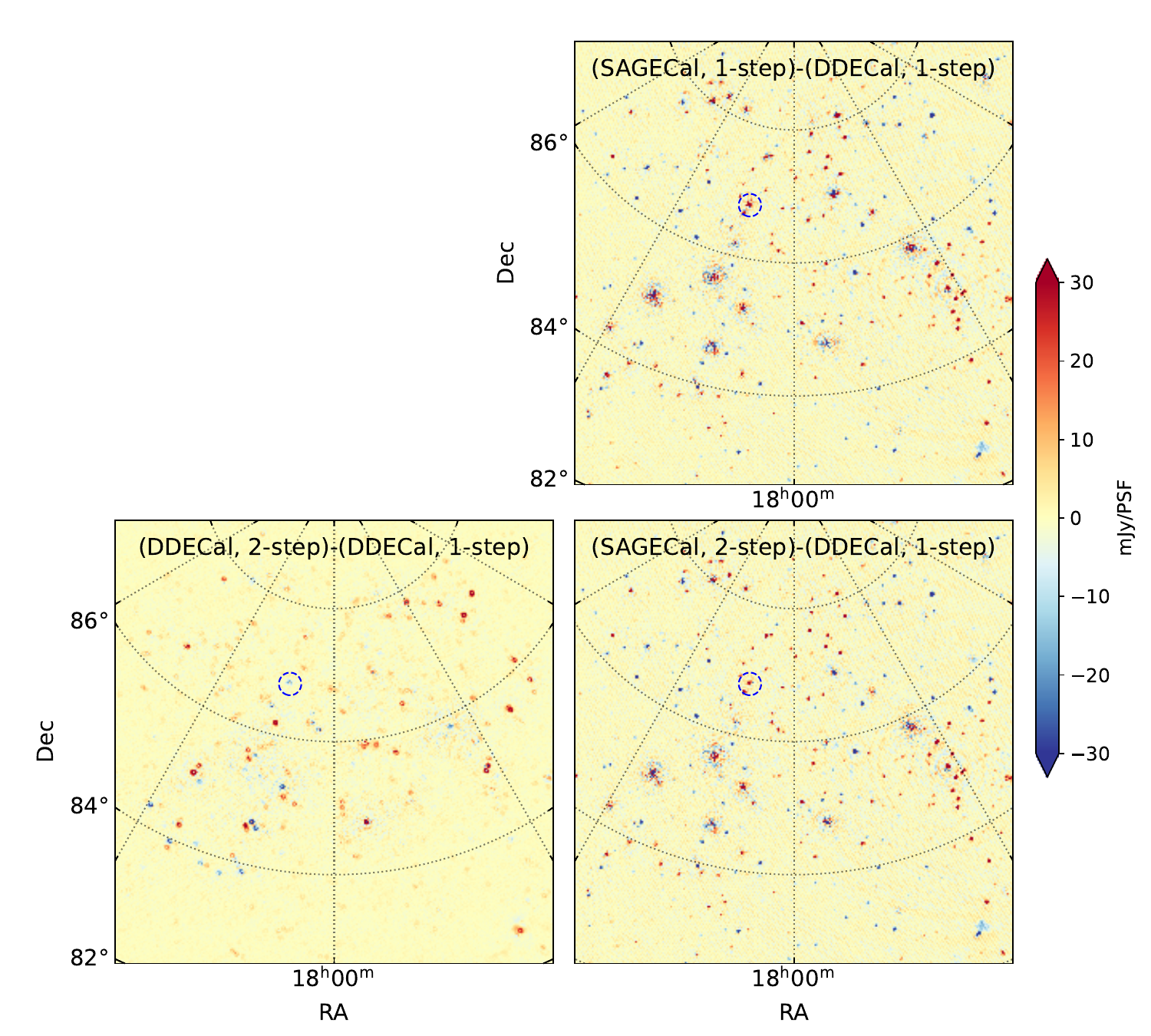}
    \caption{Difference of LOFAR-HBA $5^\circ\times5^\circ$ Stokes-I residual images after DD-calibration with different calibration scenarios on the RA 18h flanking field at frequency 113.9-127.1~MHz shown in Fig.~\ref{fig:DD_images}. The images are created with a pixel size of 0.2~arcmin using baselines $50-5000\lambda$, combining 69 sub-bands and a single observation night L612832 ($\sim11.6$-hour). The residual images with different DD-calibration scenarios are subtracted by the residual image of the \textsc{ddecal} and 1-step scenario. We mark a reference source with a dashed blue circle which is identical to the one in Fig.~\ref{fig:DD_images} to compare different DD-calibration residuals. }
    \label{fig:diff_DD}
\end{figure*}

In Fig.~\ref{fig:diff_DD}, we show the difference in residuals after different calibration scenarios. The Stokes-I residual images after DD-calibration with the four scenarios in Fig.~\ref{fig:DD_images} are respectively subtracted by the residual image of the \textsc{ddecal} and 1-step method (middle left in Fig.~\ref{fig:DD_images}).

In Fig.~\ref{fig:diff_DD}, red indicates under-subtraction and blue indicates over-subtraction of sources, compared to the DD-calibration scenario with \textsc{ddecal} and 1-step method. We use the same reference source (marked with a dashed blue circle) to compare residuals after DD-calibration. Residuals of the source appear lightly blue (indicating marginal over-subtraction) in the \textsc{ddecal} and 2-step method scenario (bottom left) while residuals appear red in the two \textsc{sagecal} scenarios (top and bottom right), indicating under-subtraction.

Overall, \textsc{ddecal} with the 2-step method shows under-subtraction with most sources appearing red and relatively compact. \textsc{sagecal} shows rather scattered residuals with a mixture of over-subtraction and under-subtraction for sources. The difference between the 1-step and 2-step methods is rather small for \textsc{sagecal}.

\section{NCP sky images after DI- and DD-calibration}
\label{app:diff_images}
\begin{figure*}
    \centering
    \includegraphics[width=\textwidth]{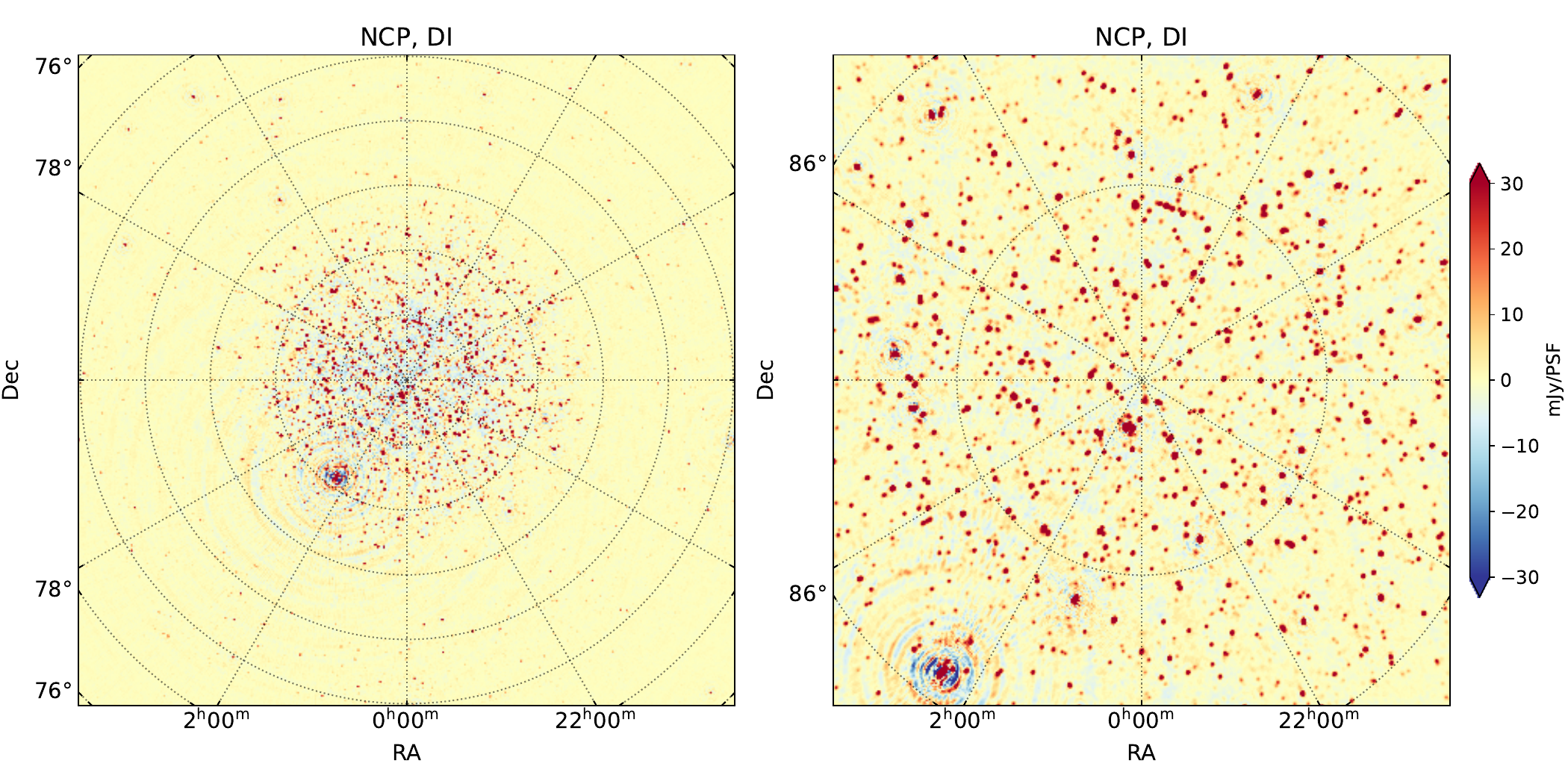}
    \includegraphics[width=\textwidth]{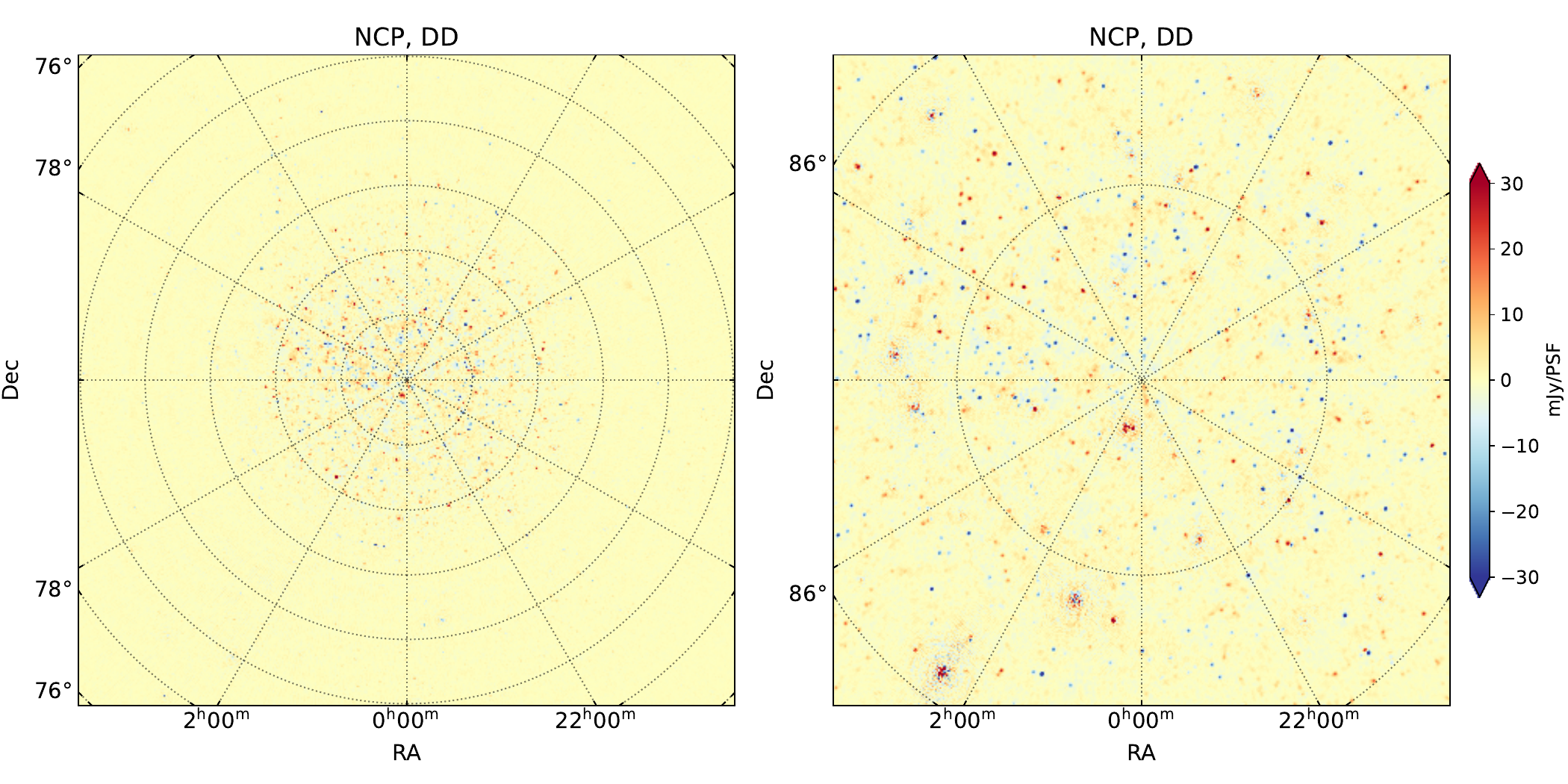}
    \caption{LOFAR-HBA Stokes-I images of the NCP at frequency 113.9 - 127.1 MHz. All 69 sub-bands are combined for imaging after the DI-calibration (in the top row) and DD-calibration (in the bottom row). The DI and DD-calibrations are performed by \textsc{sagecal} and the images are deconvolved by \textsc{wsclean}. \textbf{Left column:} A $20^\circ\times20^\circ$ image with a pixel size of of $0.8\text{ arcmin}$ with baselines between 50-1000$\lambda$ after DI-calibration (on top) and after DD-calibration (on bottom). \textbf{Right column:} A zoomed $4^\circ\times4^\circ$ image with a pixel size of of $0.2\text{ arcmin}$ with baselines between 50-5000$\lambda$ after DI-calibration (on top) and after DD-calibration (on bottom).}
    \label{fig:NCP_DI_DD_images}
\end{figure*}

In Fig.~\ref{fig:NCP_DI_DD_images}, we present the $20^\circ\times20^\circ$ and zoomed ($4^\circ\times4^\circ$) images on the NCP after DI-calibration (on top) and after DD-calibration (on bottom). Compared to the RA 18h flanking field results in Fig.~\ref{fig:DD_images}, the NCP shows better subtraction of the foregrounds in the primary beam region, these are likely due to the application of an extended sky model ($\sim8$ times more components compared to the RA 18h flanking field sky model) and more directions ($\sim6$ times more directions compared to the RA 18h flanking field) during the DD-calibration. The downside is that using an extended sky model and solving gains for more direction can be computationally much more expensive (the NCP standard processing time is $\sim5$ times longer compared to the flanking field processing time, in this work). In this work, we note that power spectra similar to the ones obtained by the NCP processing can be achieved with a relatively simple sky model and fewer directions during the DD-calibration, if the GPR foreground removal technique is applied.

\section{Gain dynamic spectra}
\label{app:gains}

\begin{figure*}
    \centering
    \includegraphics[width=\textwidth]{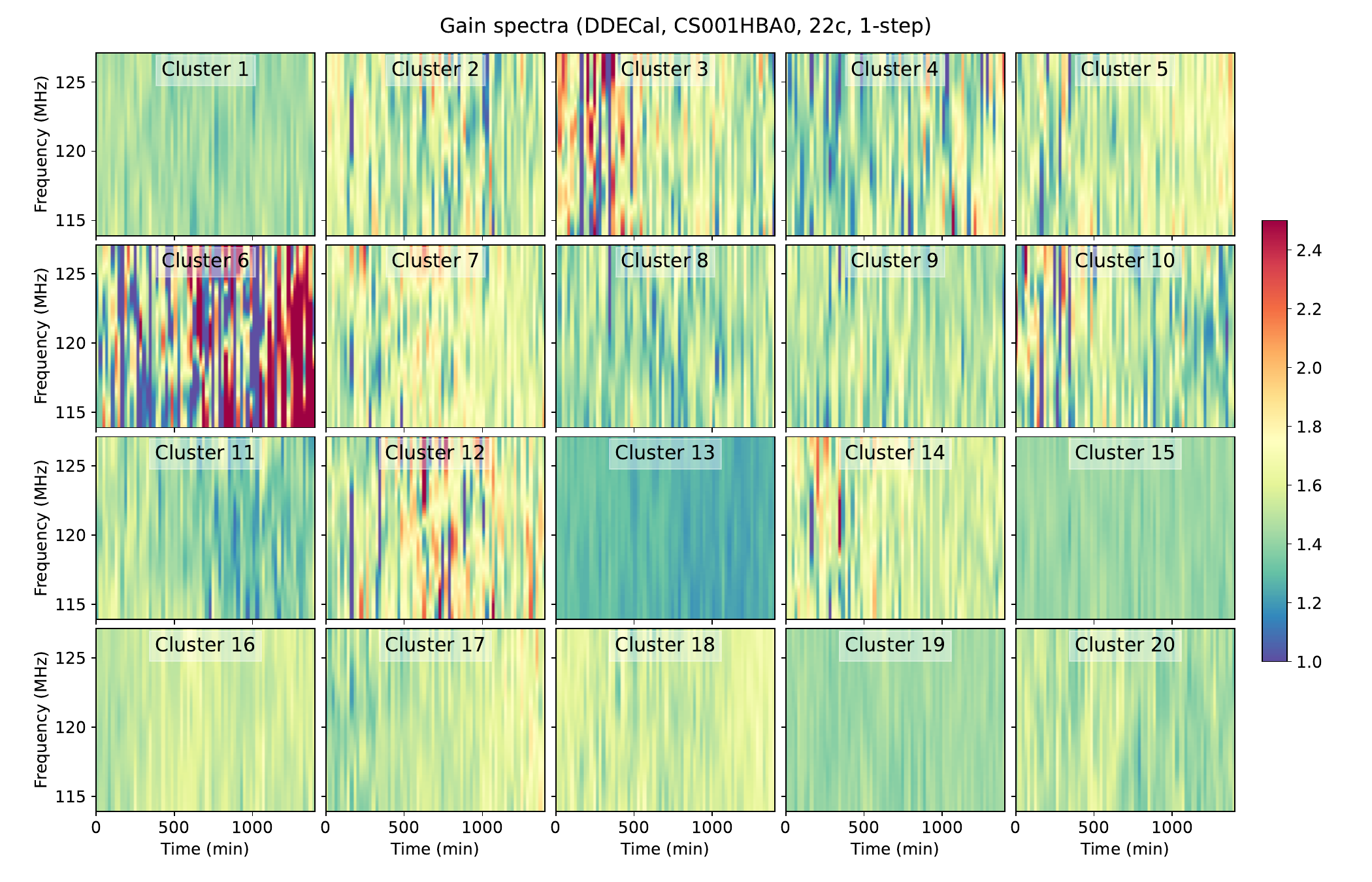}
    \includegraphics[width=\textwidth]{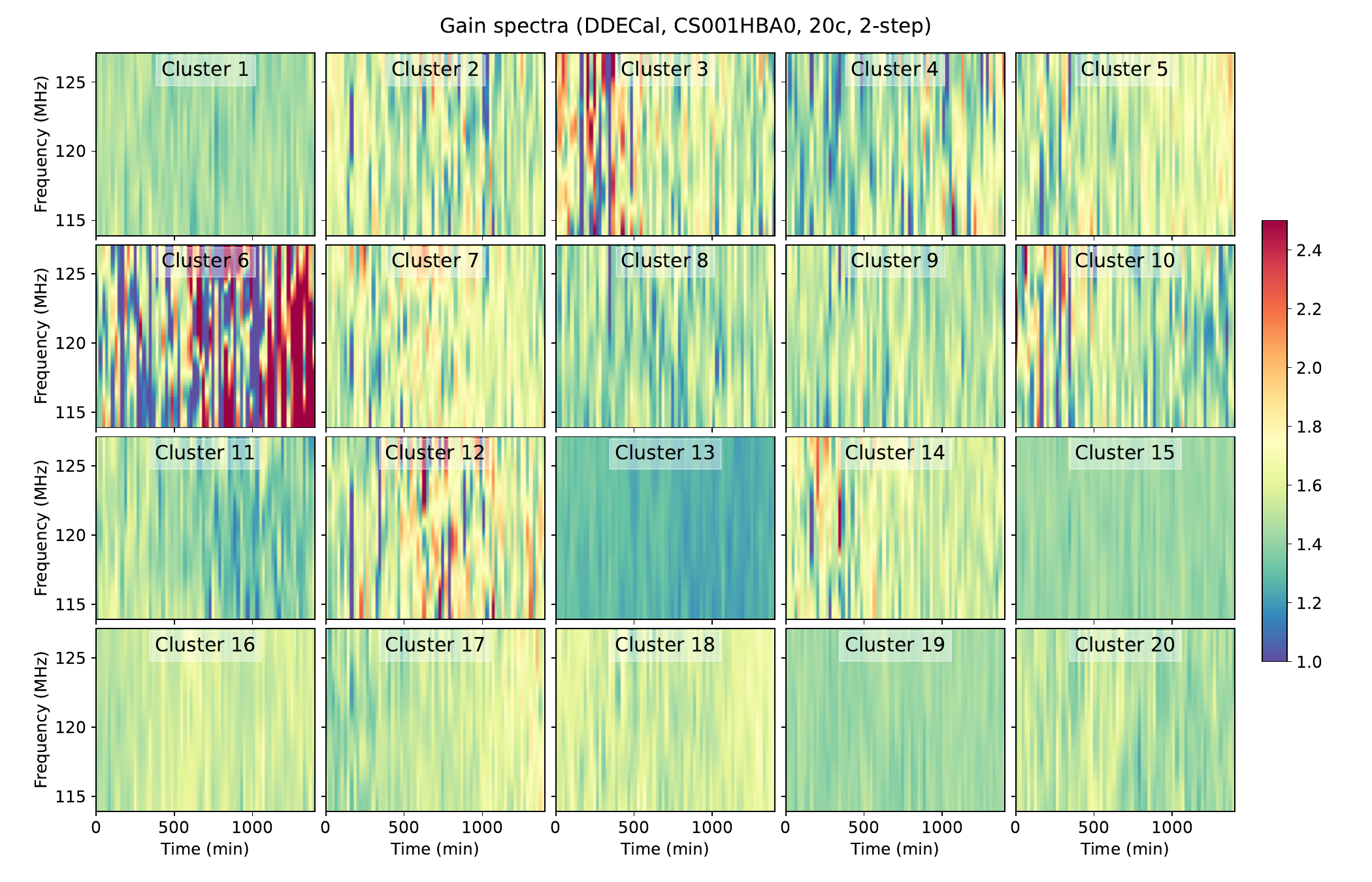}
    \caption{Gain dynamic spectra obtained by \textsc{ddecal} algorithm for 20 clusters around the phase centre, using the 1-step method (on top) and 2-step method (on bottom) for one station (CS001HBA0). Different polarisation components are added in quadrature.}
    \label{fig:gain_spectra_ddecal}
\end{figure*}

\begin{figure*}
    \centering
    \includegraphics[width=\textwidth]{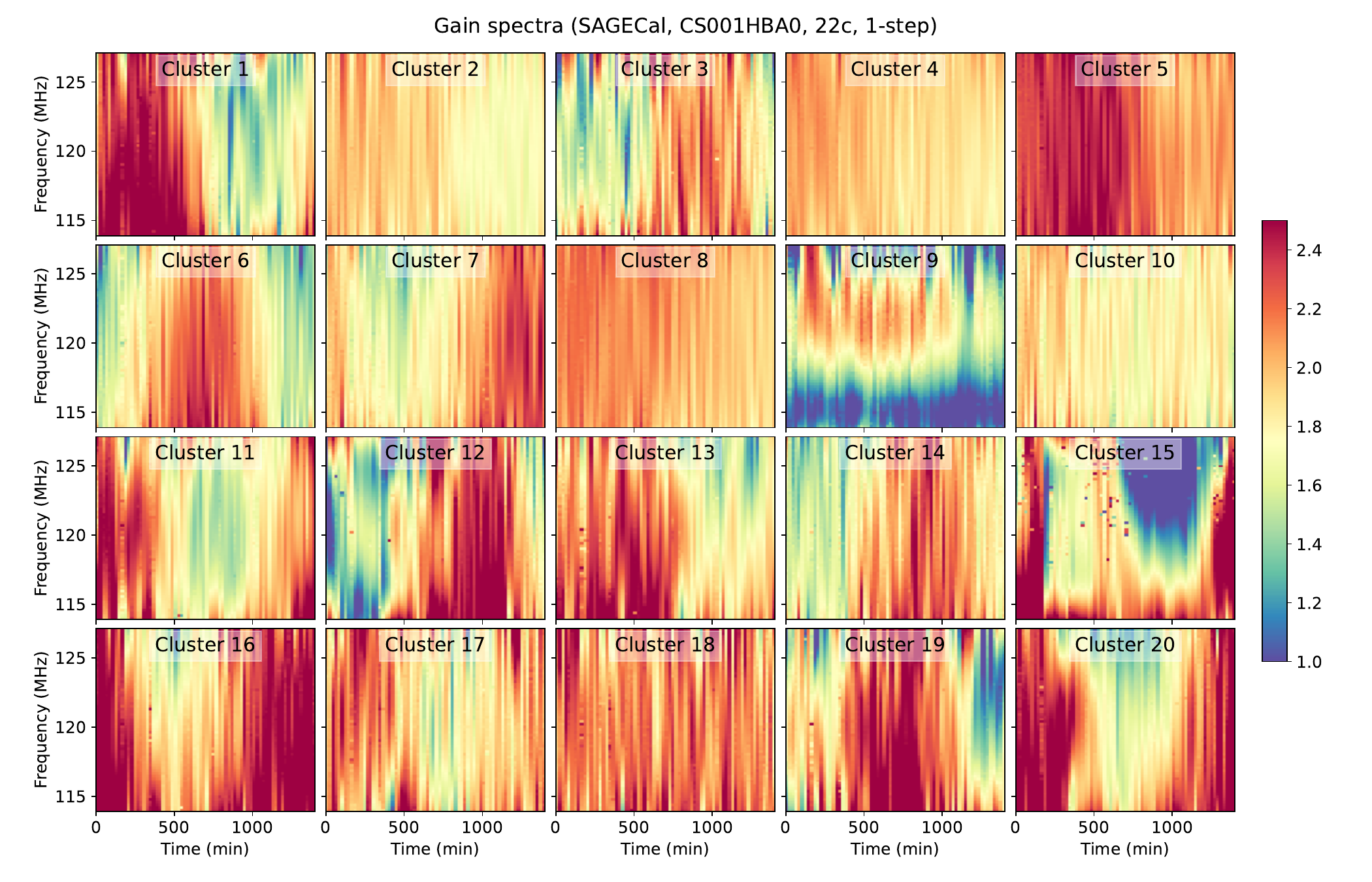}
    \includegraphics[width=\textwidth]{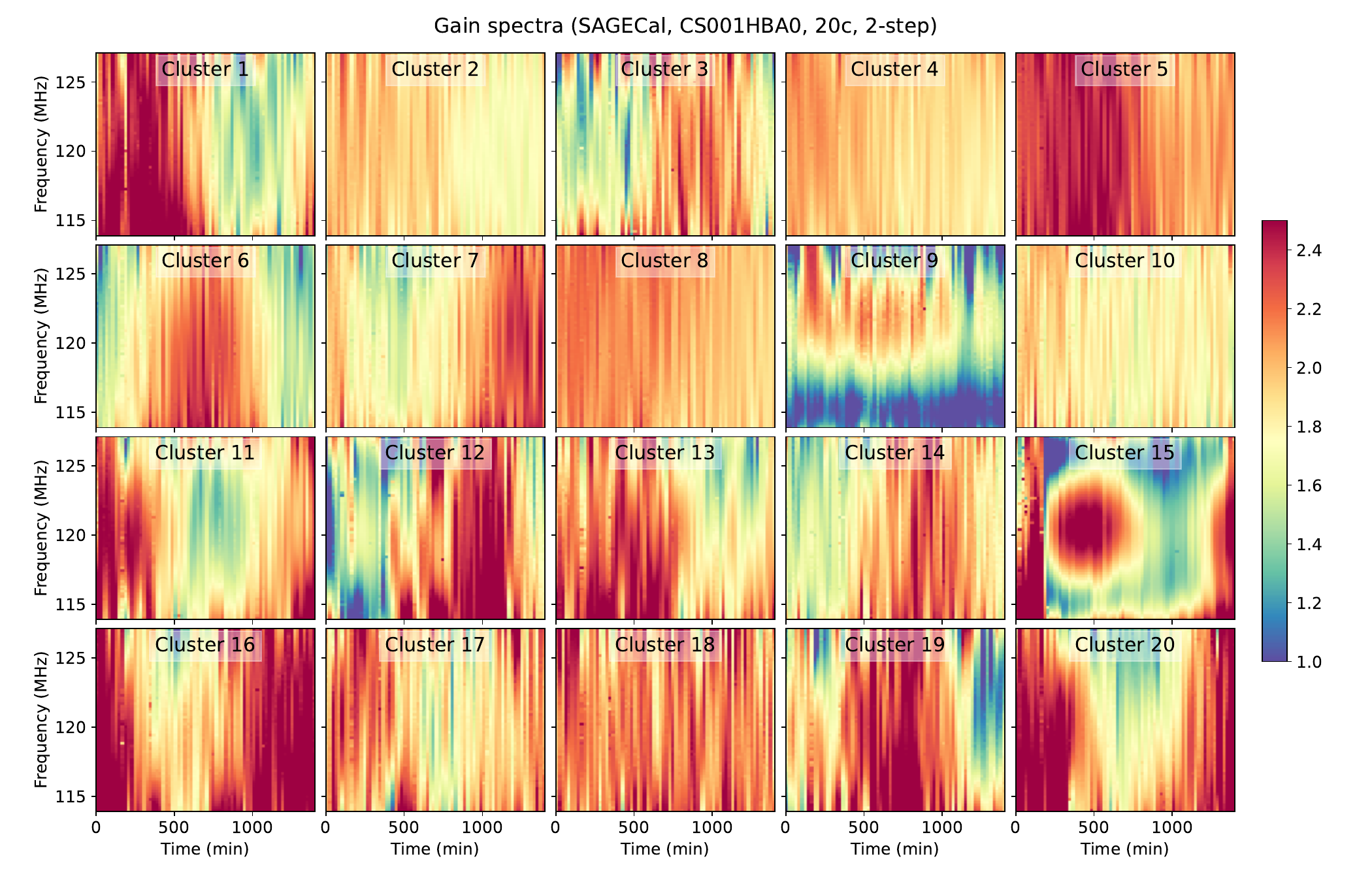}
    \caption{Gain dynamic spectra obtained by \textsc{sagecal} algorithm for 20 clusters around the phase centre, using the 1-step method (on top) and 2-step method (on bottom) for one station (CS001HBA0). Different polarisation components are added in quadrature.}
    \label{fig:gain_spectra_sagecal}
\end{figure*}

\begin{figure*}
    \centering
    \includegraphics[width=\textwidth]{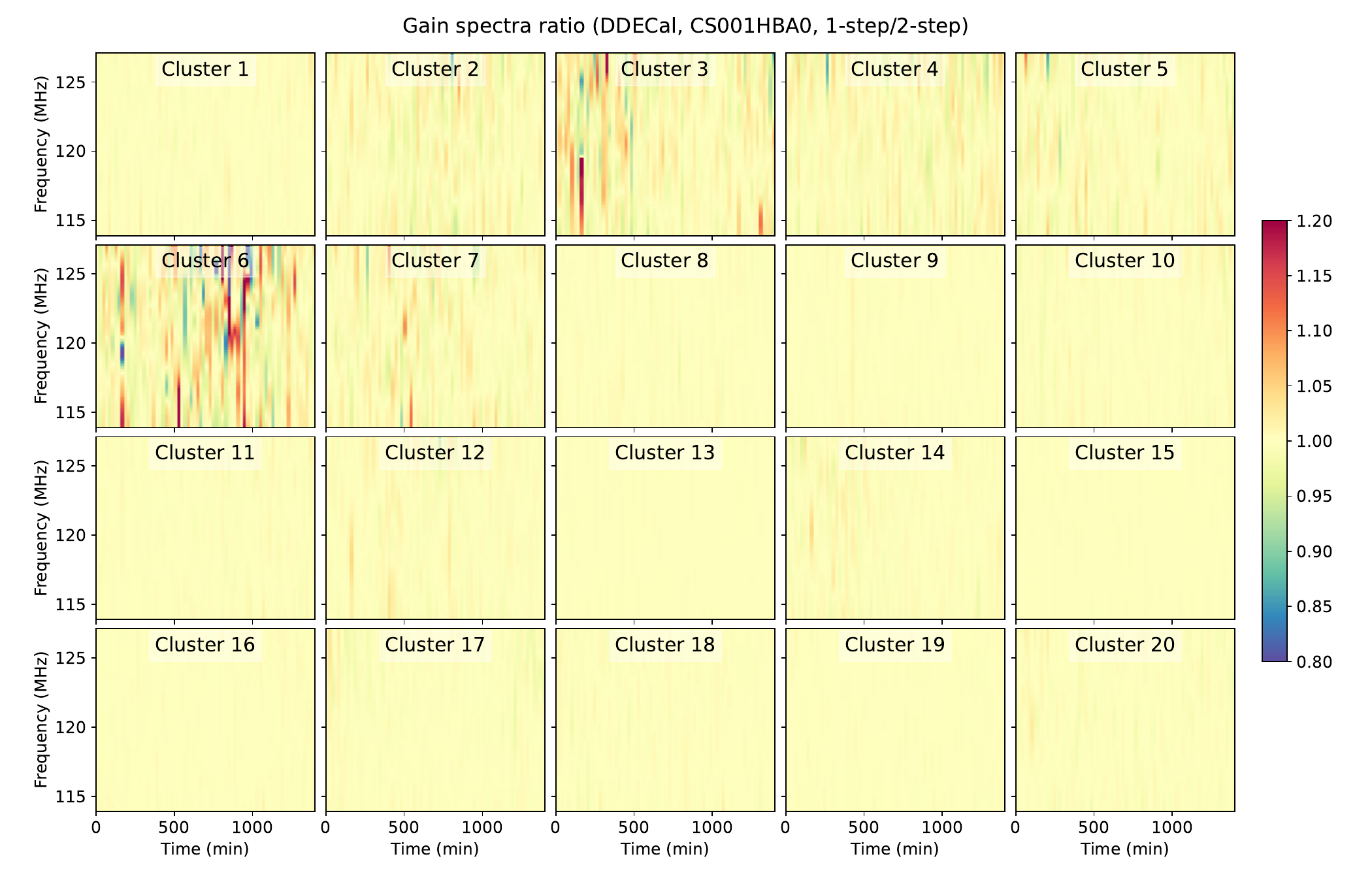}
    \includegraphics[width=\textwidth]{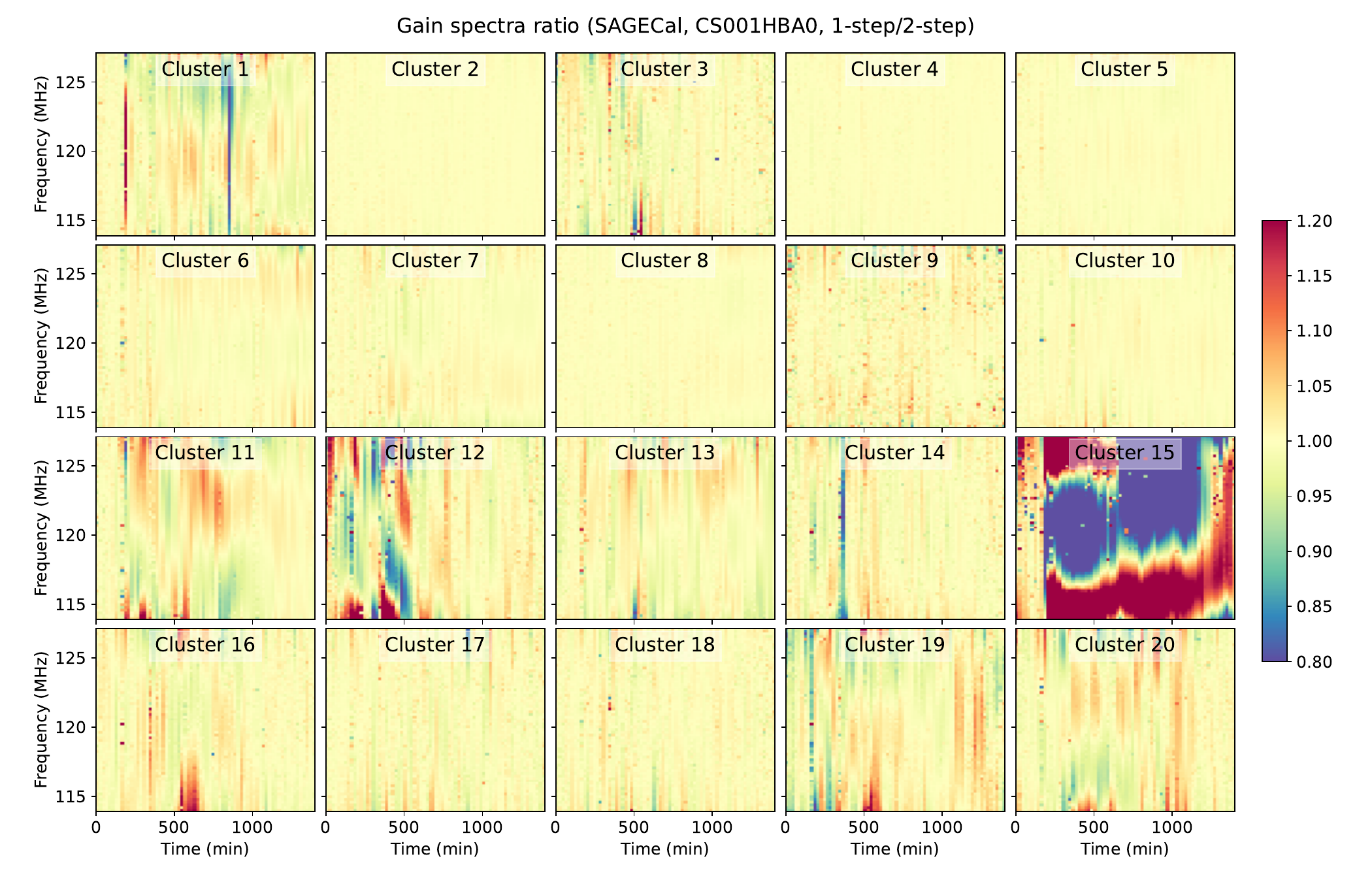}
    \caption{Ratio of gain dynamic spectra between the 1-step and 2-step methods for 20 clusters around the phase centre for one station (CS001HBA0), obtained by \textsc{ddecal} algorithm (on top) and \textsc{sagecal} algorithm (on bottom). Different polarisation components are added in quadrature. While gain differences between the 1-step and 2-step methods are more prominent in the phase centre for \textsc{ddecal} (in cluster 2-7 on top), the difference is more obvious outside the phase centre for \textsc{sagecal} (in cluster 11-20 on bottom). }
    \label{fig:gain_ratio}
\end{figure*}

\begin{figure}
    \centering
    \includegraphics[width=\columnwidth]{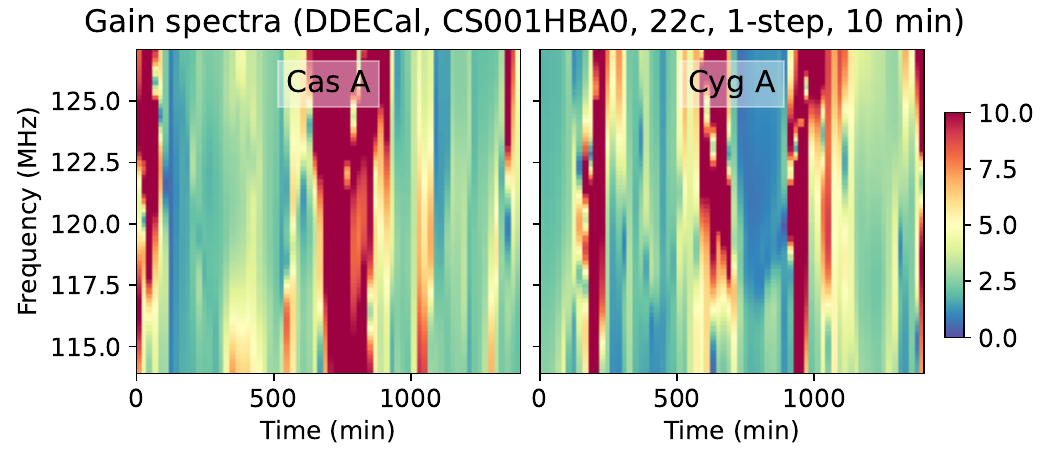}
    \includegraphics[width=\columnwidth]{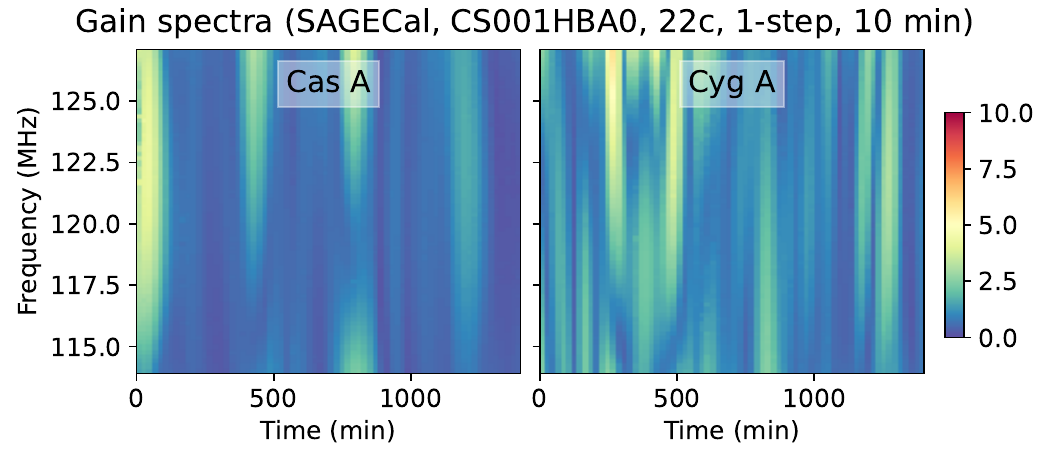}
    \includegraphics[width=\columnwidth]{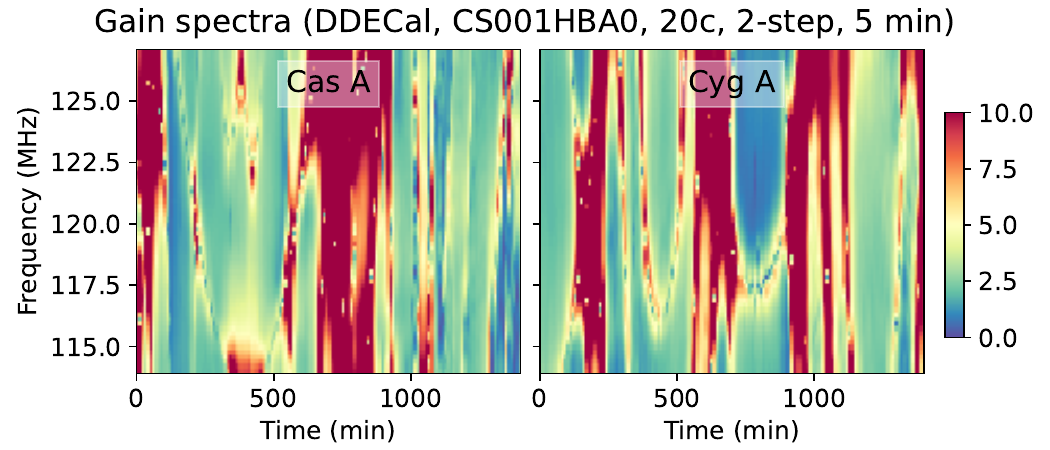}
    \caption{Gain spectra of Cas~A (left) and Cyg~A (right) obtained by different calibration strategies for one station (CS001HBA0). Different polarisation components are added in quadrature.}
    \label{fig:cc_gain_spectra}
\end{figure}

Here, we present the gain spectra of one station (CS001HBA0) per cluster obtained with the 1-step (on top) and 2-step (on bottom), achieved by the calibration algorithm, either \textsc{ddecal} (in Fig.~\ref{fig:gain_spectra_ddecal}) or \textsc{sagecal} (Fig.~\ref{fig:gain_spectra_sagecal}). From each figure, by comparing the gain spectra between the 1-step and 2-step methods, we can find how the subtraction of Cas~A and Cyg~A impacts the gains of the remaining sources in the phase centre.

Depending on the flux type of the sky model used for the calibration, obtained gains show distinct values. \textsc{ddecal} uses an intrinsic sky model and \textsc{sagecal} uses an apparent model, average gain values are higher for \textsc{sagecal} than \textsc{ddecal}. The gain spectra of \textsc{ddecal} are also flatter than \textsc{sagecal}, due to the application of the beam model.

To investigate the difference between the 1-step and 2-step methods, we create gain ratio spectra between the two methods given a calibration algorithm in Fig.~\ref{fig:gain_ratio} for CS001HBA0 per cluster. We find that the gain difference between the 1-step and 2-step methods is rather big in \textsc{sagecal} than in \textsc{ddecal}. While this difference is more concentrated in the clusters close to the phase centre for \textsc{ddecal}, from cluster 2 to 7, the difference is more obvious in outer clusters, from cluster 11 to 20, for \textsc{sagecal}.

We also present the Cas~A and Cyg~A gain spectra from different calibration setups in Fig.~\ref{fig:cc_gain_spectra}. First, we compare the gain spectra obtained by the 1-step method using \textsc{ddecal} (in the first row) and \textsc{sagecal} (in the second row). The solution interval is 10 min for both cases. The gains vary more rapidly over time for \textsc{ddecal}, in this case, and gains from \textsc{sagecal} are rather flat. In particular, the gains from \textsc{sagecal} are smoother in frequency, compared to \textsc{ddecal}. This improved smoothness in frequency of \textsc{sagecal} possibly contributed to the better performance of the subtraction of Cas~A and Cyg~A in Fig.~\ref{fig:very_wide_sap005_dd}.

In the last row of Fig.~\ref{fig:cc_gain_spectra}, we present the Cas~A and Cyg~A gain spectra obtained by \textsc{ddecal} and the 2-step method. The solutions have a higher resolution, i.e., a 5 min interval, and the gain spectra have more structures in time, compared to the lower resolution solutions in the first two rows. While \textsc{sagecal} with the 1-step method (in the second row) and \textsc{ddecal} with the 2-step method (in the last row) show comparable performance in subtracting Cas~A and Cyg~A, it is yet unclear whether the added structures in the gain spectra obtained by \textsc{ddecal} and the 2-step method are physical or noise. And more studies are needed to find out optimal frequency constraints and solution time intervals for calibration.

\end{appendix}

\end{document}